\documentclass[pdflatex,sn-standardnature]{sn-jnl}
\usepackage{times}
\usepackage{gensymb}
\usepackage{natbib}

\usepackage{ulem}
\jyear{2023}%

\usepackage{etoolbox}

\newcommand{\Ni}{^{56}{\rm Ni}}
\newcommand{\Qnuc}{Q_{\rm nuc}}
\newcommand{\Qe}{Q_{\rm eng}}
\newcommand{\QTn}{QT_{\rm nuc}}
\newcommand{\QTe}{QT_{\rm eng}}
\newcommand{\LTm}{LT_{\rm {-{nuc}}}}
\newcommand{\Ee}{E_{\rm eng}}
\newcommand{\te}{t_{\rm eng}}

\everycr={\noalign{\global\advance\stepno by 1}}%

\newcommand\aap{Astron. Astrophys.}                
\newcommand\aj{Astron. J.}                   
\newcommand\apj{Astrophys. J.}                 
\newcommand\apjl{Astrophys. J.}                
\newcommand\apjs{Astrophys. J. Supp.}               
\newcommand\mnras{Mon. Not. R. Astron. Soc.}             
\newcommand\pasp{PASP}               

\begin{document}
\title[Engines in stripped-envelope supernovae]{Stripped-envelope supernova light curves argue for central engine activity}
\author[1]{\fnm{\'{O}smar} \sur{Rodr\'{i}guez}}
\author*[1]{\fnm{Ehud} \sur{Nakar}}\email{udini@tauex.tau.ac.il}
\author[1]{\fnm{Dan} \sur{Maoz}}
\affil[1]{School of Physics and Astronomy, Tel-Aviv University, Tel-Aviv, Israel}

\abstract{
The luminosity of ``stripped-envelope supernovae'', a common type of stellar explosions, has been generally thought to be driven by the radioactive decay of the nickel synthesized in the explosion and carried in its ejecta. 
Additional possible energy sources have been previously suggested\cite{sharon2020,Ertl2020,afsariardchi2021, woosley2021,Sollerman2022}, but these claims have been statistically inconclusive or model-dependent. Here, we analyse the energy budget of a sample of 54 well-observed stripped-envelope supernovae of all sub-types, and present statistically significant, largely model-independent, observational evidence for a non-radioactive power source in most of them (and possibly in all).
We consider various energy sources, or alternatively, plausible systematic errors, that could drive this result, and conclude that the most likely option is the existence of a ``central engine'', such as a magnetar (a highly magnetic neutron star) or an accreting neutron star or black hole,
operating over hours to days after the explosion. We infer from the observations constraints on the engines, finding that if these are magnetars, then their initial magnetic fields are about $10^{15}$\,G and their initial rotation period is 1--100\,ms, implying that stripped-envelope supernovae could be the formative events of magnetars. 
}

\maketitle
The light that we observe from supernovae (SNe) has several distinct energy sources. The two main ones are: the internal energy deposited in the envelope by the shock that unbinds the star, known as cooling-envelope emission; and the radioactive decay of $\Ni$ that is synthesized during the explosion\cite{woosley2002}. Two additional
sources of energy are interaction of the ejecta with the circumstellar medium (CSM), and late energy injection by a central source -- a ``central engine''. 
Interaction with a CSM  can typically be identified by the presence of narrow lines in the SN spectrum. Central engines, however, are much harder to pin down.

Among the above energy sources, radioactive heating is considered to be the main one in SNe with compact stellar progenitors, such as the core-collapse of massive stars that have been stripped of most or all of their hydrogen envelopes at earlier stages of their stellar evolution---stripped-envelope (SE) SNe---which are sub-classified based on their spectra into types IIb, Ib, and Ic. Traditionally, a central engine has been invoked only in extremely energetic SNe, e.g. in some specific events\cite{Maeda2007,wang2017a,wang2017b,wang2019}, in luminous SNe\cite{Gomez2022}, and in superluminous SNe\cite{Inserra2013,Nicholl2017,DeCia2018,Lunnan2018}. 
However, the accepted view was that the power source of the light in most of the regular SE~SNe, is completely dominated by radioactive decay. This view has been recently challenged. A method of analysis\cite{nakar2016,wygoda2019}, based on a development of the ``Katz integral'' approach\cite{katz2013}, was used in a study\cite{sharon2020} to estimate 
the $\Ni$ mass ($M_\text{Ni}$)
and the contribution of non-$\Ni$ energy sources to the light curves of a sample of SNe of various types. The study found that, in the light curves of normal SE~SNe, there are signs of an energy contribution from a source which cannot be $\Ni$, and which was attributed to cooling emission. However, the study assessed that this excess energy is comparable to the systematic errors, and therefore considered its detection tentative. A separate study\cite{afsariardchi2021} also found hints for a contribution from a non-radioactive source in SE~SNe, based on a different approach. First, they measured $M_\text{Ni}$ in a sample of such SNe, based on
the luminosity in their radioactive tails.
They then compared the luminosity of the peak in these SE~SNe to the one predicted by numerical simulations, discovering that the observed peak luminosity 
is systematically brighter than the one predicted by simulations that are based on the measured values of $M_\text{Ni}$ . A number of additional works, modeling the progenitors and their explosions, have also suggested that the amount of $\Ni$ that can be synthesised during the explosion is insufficient to explain the peak luminosity of some observed SE~SNe\cite{Ertl2020,woosley2021,Sollerman2022}. The indications of a non-radioactive source in all but the first (and statistically inconclusive) of these previous studies are, by nature, strongly model dependent. 

We study a sample of 54 well-observed SE~SNe (\ref{fig:tL_tQ} and \ref{table:SN_sample}) from among the 191 such events analyzed in a previous work\cite{rodriguez2023}. The sample has bolometric light curves that are well-sampled over time, accurately measured $\Ni$ masses (based on the 
radioactive tail
luminosity), and relatively well constrained explosion times (see Methods for details). We use this sample to measure the contribution of non-radioactive power sources and to place constraints on the physical properties of such sources,
utilizing and extending the Katz integral\cite{katz2013} method, to separate between the $\Ni$ and non-radioactive contributions\cite{nakar2016}. In the Methods section, we briefly repeat the derivation of the technique, and extend it to account for possible energy deposition by a central engine.  

At times since explosion $t$, significantly longer than the diffusion time through the ejecta, $t_{\rm diff}$, 
the extended Katz integral (equation~(\ref{eq:tE}) in Methods) can be written as 
\begin{equation}\label{eq:ET}
\LTm \equiv LT(t)-\QTn(t)  \approx ET + \QTe(t)~~~;~~~ t \gg t_\text{diff}~.
\end{equation}
Here, $ET\equiv E(t_0)\,t_0$ is a constant, where $E(t)$ is the internal energy trapped in the ejecta as a function of time, and $t_0$ is the earliest time at which the expansion can be regarded as homologous. 
$LT(t)$,  $\QTn(t)$ and $\QTe(t)$ are the time-weighted integrals from $t_0$ to $t$ over the bolometric luminosity, $L(t)$, the energy deposition rate from radioactive decay, $\Qnuc(t)$, and the energy deposition rate from a central engine, $\Qe(t)$, respectively (e.g. $LT(t)\equiv\int_{t_0}^{t}L(t')\,t'\,dt'$).
Equation~(\ref{eq:ET}) separates the time-weighted integrated luminosity into the three possible energy sources that can power the emission (when there is no interaction with a CSM): the cooling emission $ET$, the radioactive decay $QT_{\rm nuc}$, and the central engine $QT_{\rm eng}$.  
We stress that equation~(\ref{eq:ET}) is quite robust and it is, basically, a kind of energy conservation statement that takes adiabatic losses into account. As such, it is independent of the uncertain radiative transfer in the problem as well as the velocity and abundances  distribution of the ejecta. The only assumptions that underlie this equation are that the outflow is homologous and that the internal energy is dominated by radiation. Both of these assumptions are satisfied in all SNe that are not powered by interaction with 
a CSM,
i.e., they are expected to be  applicable to all normal SNe of types II, IIb, Ia, Ib, and Ic.

\begin{figure}
\includegraphics[width=\columnwidth]{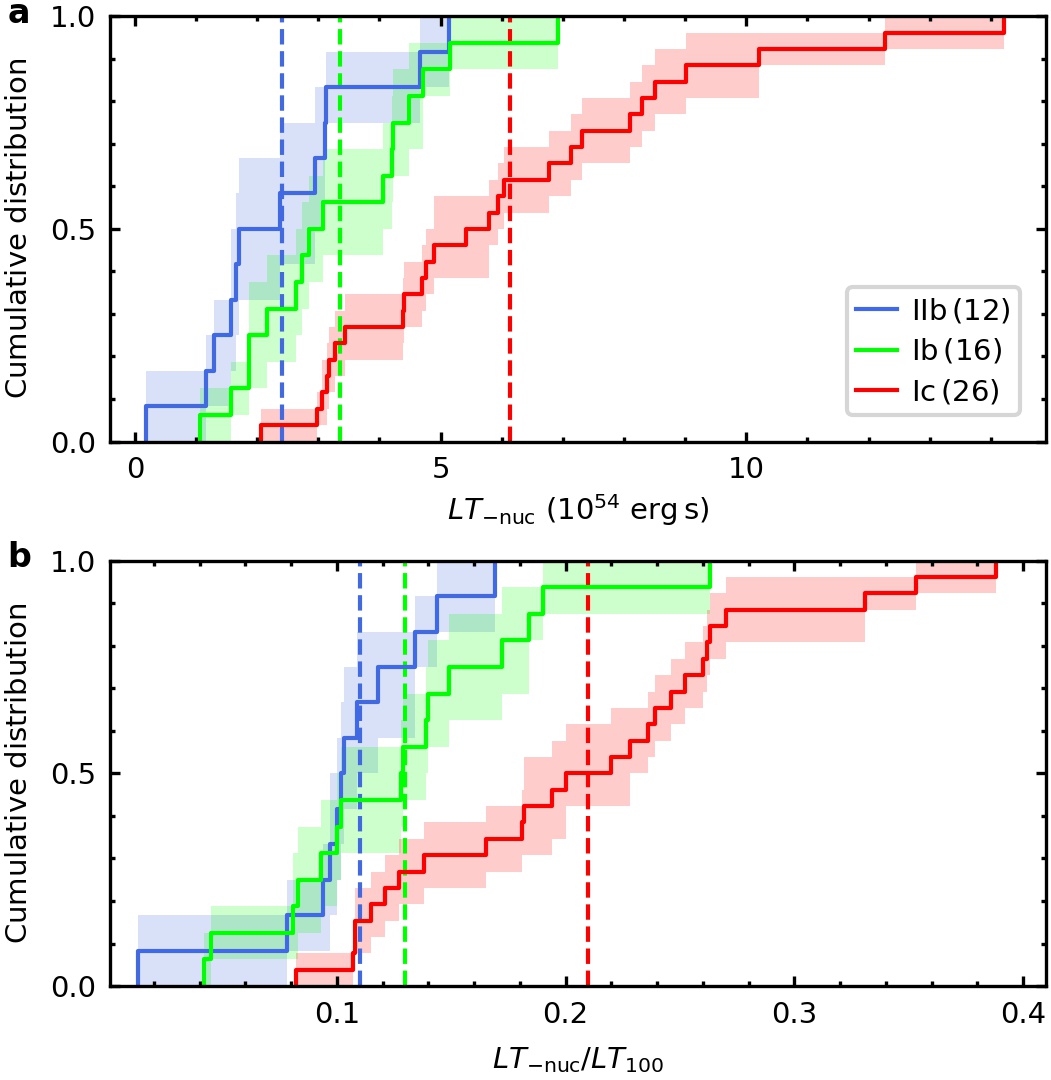}
\caption{{\bf The non-radioactive contribution to the light of SE SNe.} Cumulative distributions for $\LTm$ (\textbf{a}) and for $\LTm/LT_{100}$ (\textbf{b}). Shaded regions represent 68\% confidence intervals computed by bootstrap resampling (10,000 samples). Vertical dashed blue, green, and red lines indicate mean values for SNe~IIb, Ib, and Ic, respectively. Numbers in parentheses are the sample sizes for each SE~SN subtype.}
\label{fig:LT-nuc_cdf}
\end{figure}

We have measured the above-defined excess over the radioactive heating contribution, $\LTm$, and $\LTm/LT_{100}$, for all of the SNe in our sample 
(\ref{table:LT-nuc})
where $LT_{100} \equiv LT(t=100~{\rm d})$. Fig.~\ref{fig:LT-nuc_cdf} shows the cumulative distributions for the $\LTm$ and $\LTm/LT_{100}$ values. 
The most likely values of $\LTm$ in our sample are all greater than zero. For all SNe, except four, $\LTm$ is positive at $>2\,\sigma$ significance. The mean $\LTm$ values for the type-based subsamples of SNe~IIb, Ib, and Ic 
are $2.4\pm0.4$, $3.4\pm0.4$, and $6.1\pm0.6\times10^{54}$\,erg\,s (errors are the standard error of the mean, SEM), respectively, which
are at least 5.7\,SEM greater than zero, while the average $\LTm/LT_{100}$ value is $0.11\pm0.01$, $0.13\pm0.01$, and $0.21\pm0.02$ ($\pm1$\,SEM) for each SN type, respectively. In other words, $\LTm$ is not only significantly greater than zero---it is also a sizeable fraction of $LT_{100}$. In \ref{fig:tL_tQ} 
we plot $L(t) \cdot t$ and $Q_{\rm nuc}(t) \cdot t$ for each of the SNe in the sample. One sees that $L(t) \cdot t \approx Q_{\rm nuc}(t) \cdot t$ is reached well before our somewhat-arbitrary choice of 100\,days, and hence $\LTm/LT_{100}$ provides a conservative estimate of the non-radioactive contribution to the peak of the light curve. Our results are consistent with those of the previous study\cite{sharon2020} that  found positive (but barely significant, as assessed by the authors) values of $\LTm$ for 9 out of a sample of  11 SE~SNe . Our results are more significant owing to our larger sample (54 versus 11~SNe), in which we have carefully considered all possible sources of errors (see Methods). We note that, in that previous study\cite{sharon2020}, the authors did not consider the possibility of a central engine, and therefore $ET$ in their paper is equivalent to $\LTm$.

The energy excess we measure in SE~SNe is surprising, in view of the common wisdom that SE~SNe, after the first day post-explosion, are powered solely by radioactive decay\cite{dessart2011}. We therefore first consider the possibility that, rather than being physical, our result is an artifact of systematic errors. The observable $\LTm/LT_{100}$ is independent of the adopted distances and bolometric-correction zero points, ruling out systematic errors in these parameters as driving the results (details of the derivation of $L$ and $M_{\rm Ni}$ have been previously reported\cite{rodriguez2023}). We have further considered the possibility of systematic errors in the adopted explosion time or the assumed dust reddening correction of each SN event, in the UV and IR bolometric corrections, or in the assumption of isotropic emission, finding it highly unlikely that systematic errors in any of these can be the source of the positive values of $\LTm$ (see Methods for details). We have also contemplated the effect of the adopted deposition function of the radioactive heat. For a theoretically plausible range of deposition functions, $\LTm$ remains positive at high significance, while the $\LTm$ values vary by tens of per cents between different deposition functions (see Methods). Thus, unless the deposition function is radically different than predicted by theory, its choice is not at the root of the positive values of $\LTm$.
Moreover, many of the systematic errors would likely affect all of the SNe in our sample alike, regardless of their spectroscopic classification, but the different SN types display significantly different $\LTm/LT_{100}$ distributions. Finally, we note the previous finding\cite{sharon2020} that $\LTm$ is consistent with zero for Type Ia SNe (as expected in the thermonuclear combustion of a white dwarf, which leaves no remnant), and therefore a systematic error in our method would somehow have to be restricted to SE~SNe.

As the positive value of $\LTm$ is most likely of physical origin, and there is no evidence for a CSM interaction in our SN sample (see Methods), the remaining possible excess energy sources are either cooling emission or a central engine. The cooling-emission option is arguably inconsistent with both theory and observations. First, if cooling emission dominates, then $ET \approx \LTm \sim 10^{54}{-}10^{55}$\,erg\,s. This implies that, for the SNe in our sample, the progenitor radius is $\sim 10^{13}$\,cm, and that most of the ejecta mass is roughly at this radius before the explosion (i.e., this is not just the radius of some low-mass extended envelope; see Methods and ref.\cite{shussman2016}). Such large pre-explosion radii run counter to stellar evolution models of SE~SNe progenitors\cite{yoon2010,laplace2020}--- see Methods for a discussion. Second, a large progenitor radius sets a lower limit on the bolometric luminosity at early times. However, for almost all of the SNe in our sample, the observed upper limit on $L$ at these early times is below this minimum (see Methods). We conclude that the most likely explanation for the measured values of $\LTm$ in our sample is the activity of a central engine.

If a central engine is the dominant energy source of $\LTm$, then we can constrain some of its properties.
First, if all the injected energy thermalizes, then $\Ee \te \approx \LTm$ or, using fiducial numbers,
\begin{equation}\label{eq:Eeng_teng}
    E_{\rm eng} \approx 10^{49}\,{\rm erg}\,\left(\frac{t_{\rm eng}}{10^5\,{\rm s}}\right)^{-1}\,\frac{\LTm}{10^{54}\,{\rm erg\,s}},
\end{equation}
where, in the general case, $E_{\rm eng}$ is the energy deposited by the engine, and $\te$ is the duration of this energy injection (see Methods of the exact definitions of $E_{\rm eng}$ and $\te$, and for a discussion of cases where they differ from the definitions above).  
If not all of the energy thermalizes, then 
equation~(\ref{eq:Eeng_teng})
will be a lower limit on $\Ee$. From a combination of 
equation~(\ref{eq:Eeng_teng}) 
with the requirement that the energy is deposited in time to affect the observations, we obtain the following constraints (see Methods for details):
\begin{equation}\label{eq:eng_costrains}
   \begin{array}{c}
   10^{48}{-}10^{49}\,{\rm erg} < \Ee<10^{51}{-}10^{52}\,{\rm erg}, \\
   \\
   10^3{-}10^4\,{\rm s} < \te <10^6\,{\rm s}.
   \end{array}
\end{equation}  

Long-lasting accretion onto the compact remnant is one possible source of the engine energy, although, it may be hard for such an engine to satisfy the above constraints (see Methods). A second option, which has been previously discussed in the context of SE~SNe\cite{Woosley2010,Kasen2010}, is a newly born magnetar.
As we show here, it can naturally explain the observed $\LTm$. A magnetar taps a stellar remnant's rotational energy and injects it into the ejecta via a magnetized wind. Requiring that the rotational energy and spin-down rate of the magnetar satisfy
equations~(\ref{eq:Eeng_teng}) and (\ref{eq:eng_costrains}), we can
constrain its magnetic field as follows (see Methods).  
\begin{equation}\label{eq:Bmag}
    B \approx 10^{15}\,{\rm G} \left(\frac{\LTm}{4 \times 10^{54}\,{\rm erg\, s} }\frac{\ln(t_{\rm peak}/t_{\rm mag})}{5}\right)^{-1/2},
\end{equation}
where $B$ is the surface dipolar magnetic field, $t_{\rm mag}$ is the magnetar energy deposition time, and $t_{\rm peak}$ is the time where the observed luminosity peaks. The initial rotation period is constrained to the range $\sim 1-100$\,ms (see Methods).
These values for the magnetic field and the initial period align well with those expected for magnetars at birth.

SE~SNe constitute $\sim30\%$ of all core-collapse SN explosions\citep{shivvers2017}, and so their roles are comparably important to other SN types in cosmic element synthesis and diffusion, in energy feedback in star and galaxy formation, in cosmic ray acceleration within old SN remnants, and potentially in the formation of the population of stellar-mass black holes. Our measurements and analysis suggest that in SE~SNe there is a significant contribution to the observed luminosity, by post-explosion energy injection from a central object, on timescales of hours to days. 
If this central engine is a young, fast-spinning, magnetar, then its properties appear remarkably consistent with expectations for objects of this class at their formative phase. It has been estimated that 25--70\% ($\pm1\,\sigma$ range) of neutron stars are born in SNe as magnetars\citep{biniamini2019}, and hence our result is consistent with SE~SNe being the main formation events of magnetars. Nonetheless, in other types of core-collapse SNe, the cooling emission phase is much more extended ($\sim 100$\,days, a result of the large progenitor radius), and therefore the Katz integral method is much less sensitive to the presence of central engines like those we have found here for SE~SNe, and which would contribute in other SN types only of order a few per~cent excess to the energy$\times$time budget. We therefore cannot exclude that other types of core-collapse SNe can perhaps also produce magnetars. 


\newpage

\backmatter


\section*{Methods}
\counterwithin{figure}{section}
\renewcommand{\thefigure}{Extended Data Fig.\,\arabic{figure}}
\renewcommand{\figurename}{} 
\setcounter{figure}{0}
\renewcommand{\thetable}{Extended Data Table\,\arabic{table}}
\renewcommand{\tablename}{} 
\setcounter{table}{0}

\subsection*{The Katz Integral}\label{sec:katz_integral}
We briefly repeat the derivation of the Katz integral\cite{katz2013} and its extension\cite{nakar2016} to study energy sources beyond $\Ni$.
In the Main Text, we further develop the separation of the $\Ni$ and cooling-emission to take into account the contribution of a long lasting engine.
Consider a non-relativistic, homologous, expanding outflow, with internal energy as a function of time since explosion, $E(t)$, dominated by the photons that are trapped in the ejecta. The derivative of this internal energy satisfies
\begin{equation}\label{eq:dEt_dt}
dE(t)/dt = -E(t)/t + Q_\text{dep}(t)-L(t),
\end{equation}
where the first term on the r.h.s accounts for the adiabatic losses of the trapped radiation, and the last term is the luminosity of the thermalized photons that escape the ejecta. $Q_\text{dep}(t)$ is the instantaneous energy deposition rate, in the form of thermalized photons, which in the absence of interaction of the ejecta with a CSM, has only two known possible sources---radioactive decay and the activity of a central engine:
\begin{equation}\label{eq:Qdep}
Q_{\rm dep} = \Qnuc + \Qe .
\end{equation}
Rearranging this derivative and integrating over the time, one obtains the Katz integral,
\begin{equation}\label{eq:tE}
E(t)\,t-E(t_0)\,t_0=\int_{t_0}^{t} (Q_\text{dep}(t')-L(t'))\,t'\,dt'.
\end{equation}
The integration is from $t_0$, the time at which homologous expansion begins, until any later time $t$. Since, by time $t_0$, the radiation energy has become about an order of magnitude smaller than the total explosion energy, and the acceleration of the ejecta has become negligible, $E(t_0)$ is still completely dominated by the energy deposited by the shock that unbounds the stellar envelope, while the energy deposition by radioactive decay is not yet significant. 
When the time since explosion (which is also the dynamical time) becomes significantly longer than the diffusion time through the ejecta, $t_{\rm diff}$, all of the energy that was trapped in the ejecta has escaped, and the term $E(t)\,t$ becomes comparable to $Q_\text{dep}(t) \cdot t_{\rm diff}(t) \cdot t$, which vanishes with growing $t$.

\subsection*{Data Set}\label{sec:data_set}
In a recent analysis\cite{rodriguez2023}, data were collected for 191~SE~SNe from the literature and from the Zwicky Transient Facility (ZTF) Bright Transient Survey\cite{fremling2020,perley2020}, having known redshifts and photometry at peak light in at least two optical bands. For those SNe, the authors calculated distance moduli ($\mu$), host galaxy reddenings ($E(B-V)_\text{h}$), explosion epochs ($t_\text{exp}$), and luminosities ($L$). To compute the luminosities, they used 
optical photometry 
(Johnson--Kron--Cousins $BV\!RI$, Sloan $gri$, and/or ZTF $gr$) and new bolometric corrections (BC), derived using 15~SNe~IIb, 15~SNe~Ib, and 19~SNe~Ic with optical and near-IR photometry from $B$ to $H/K$ (0.44 to 1.65/2.16\,\textmu m), including corrections to account for the unobserved flux at wavelengths $<0.44$\,\textmu m and $>1.65/2.16$\,\textmu m. From the luminosity light curves, they computed peak luminosities ($L_\text{peak}$), the corresponding peak times ($t_\text{peak}$), and decline rates defined as $\Delta m_{15}=2.5\,\log(L_\text{peak}/L(t_\text{peak}+15\,\text{d}))$. They further collected ejecta velocities at $t_\text{peak}$ ($v_\text{peak}$), measured from absorption lines in optical spectra. For 75~SNe they computed $M_\text{Ni}$ and gamma-ray escape times ($t_\text{esc}$) from the radioactive tail of their luminosity light curves, assuming a form\cite{clocchiatti1997} for the deposition function.

Among the 75~SE~SNe with $M_\text{Ni}$ estimates, we select for this work 54~SNe that have their first luminosity measurements at times $t_\text{first}$ $\lesssim10$\,d since explosion. Our final sample is listed in \ref{table:SN_sample}, which includes the SN name, the spectral subtype, $\mu$, the heliocentric SN redshift $z$ (throughout this work, all observed time intervals have been corrected by a factor $1+z$ to bring them to the SN rest frame), the Galactic reddening\cite{schlafly2011} $E(B-V)_\text{G}$, $E(B-V)_\text{h}$, the absolute $r$-band magnitude at peak $M_{r,\text{peak}}$, the epochs of last non-detection $t_\text{non-det}$ and first SN detection $t_\text{detect}$, $t_\text{exp}$, and the filters of the photometry used by ref.\cite{rodriguez2023} to compute luminosities through the BC technique.
The luminosity light curves of the SNe in our sample, multiplied by the time since explosion, are shown in \ref{fig:tL_tQ}. \ref{table:SN_parameters} lists SN parameters, including $M_\text{Ni}$, $t_\text{esc}$, $t_\text{peak}$, $L_\text{peak}$,
$\Delta m_{15}$, and $v_\text{peak}$.

\subsection*{Estimating $\LTm$ from the observations}\label{sec:Calculate_LTm}
\begin{figure*}[!h]
\includegraphics[width=1.0\textwidth]{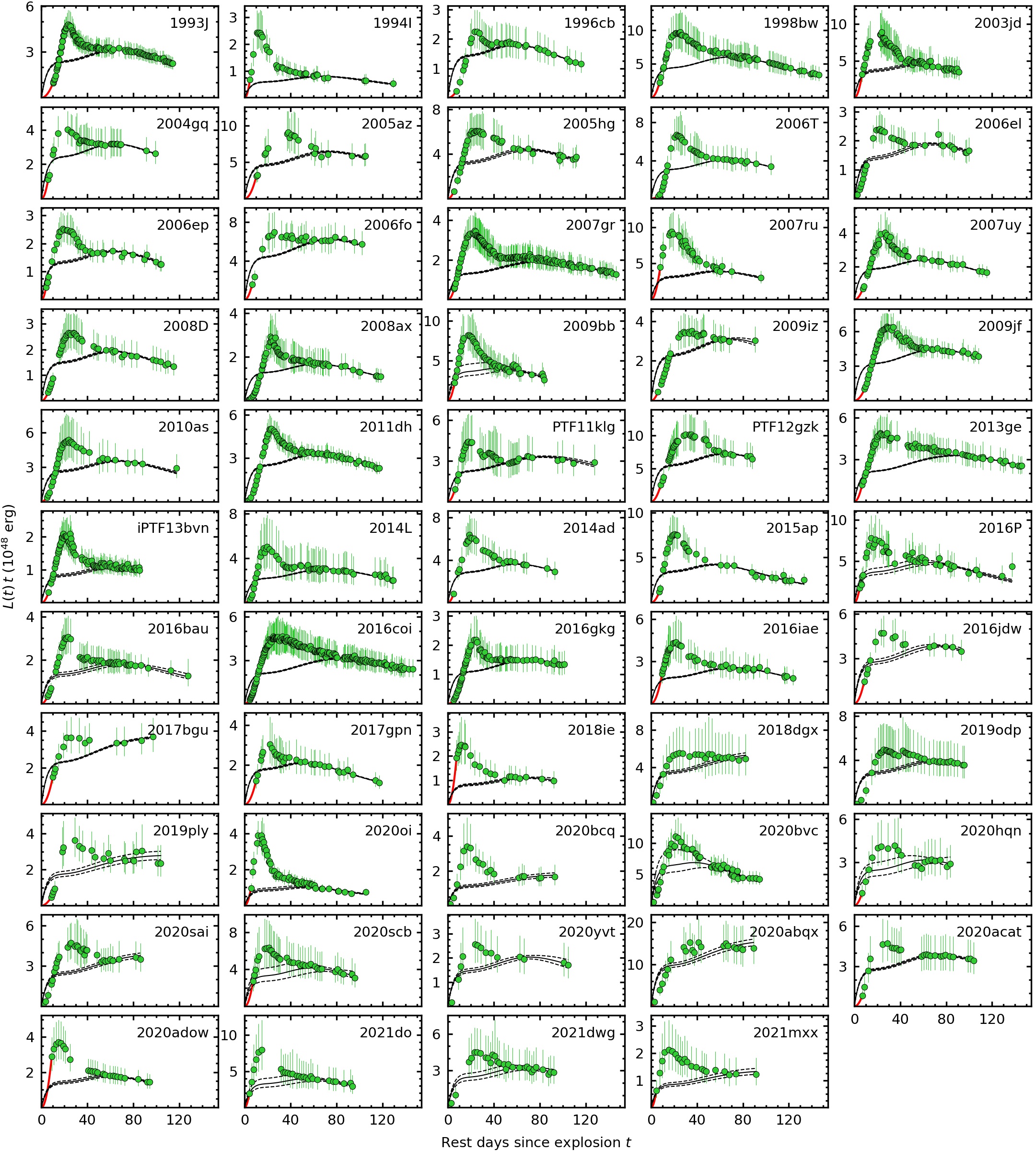}
\caption{{\bf Time-weighted luminosity light curves for the SNe in our sample.} Red solid lines are extrapolations of $L(t)\,t$ to $t=0$. Black solid lines are time-weighted energy injection due to the radioactive decay chain ${^{56}\text{Ni}}\to{^{56}\text{Co}}\to{^{56}\text{Fe}}$, while dashed lines correspond to the 5th--95th~percentile error range due to uncertainties in $M_\text{Ni}$ and $t_\text{esc}$. Error bars are $1\sigma$ and include propagated uncertainties in distance, reddening, bolometric correction, and photometry.}
\label{fig:tL_tQ}
\end{figure*}

The observable $LT(t)$ can be expressed as
\begin{equation}\label{eq:LT_last}
LT(t)=LT(t_\text{first}) + \int_{t_\text{first}}^{t}L(t')\,t'\,dt',
\end{equation}
where $LT(t_\text{first})=\int_0^{t_\text{first}}L(t')\,t'\,dt'$ is the contribution to $LT(t)$ from the unobserved luminosity at epochs earlier than $t_\text{first}$ (the difference between integrating from $t=0$ and integrating from $t=t_0$ is negligible). To estimate $LT(t_\text{first})$, we extrapolate the ($t_\text{first},\,L(t_\text{first})$) point linearly to zero luminosity at $t=0$. Thus, $LT(t_\text{first})$ is equal to $L(t_\text{first})\,t_\text{first}^2/3$, for which we assume a conservative relative error of 20\%. The extrapolations of $L(t)\,t$ between the assumed explosion times and $t_\text{first}$ are shown as red solid lines in \ref{fig:tL_tQ}.

For SE~SNe in general, the dominant radioactive decay chain is ${{^{56}\mathrm{Ni}}\to{^{56}\mathrm{Co}}\to{^{56}\mathrm{Fe}}}$, so
\begin{equation}\label{eq:QMq}
\Qnuc(t)=[q_\gamma(t)\,f_\mathrm{dep}(t,t_\text{esc})+q_\mathrm{pos}(t)]\,M_\text{Ni}/M_{\odot}.
\end{equation}
Here, $q_\gamma(t)$ and $q_\mathrm{pos}(t)$ are the total energy release rates of gamma-rays and positron kinetic energy per unit $M_\text{Ni}$, respectively, while $f_\mathrm{dep}(t,t_\text{esc})$ is the gamma-ray deposition function, which describes the fraction of the generated gamma-ray energy deposited in the ejecta. The terms $q_\gamma(t)$ and $q_\mathrm{pos}(t)$ are given\citep{wygoda2019} by
\begin{equation}
q_\gamma(t)=\left(6.45\,e^{-\frac{t}{t_\mathrm{Ni}}}+1.38\,e^{-\frac{t}{t_\mathrm{Co}}}\right)\times 10^{43}\,\mathrm{erg}\,\mathrm{s}^{-1}
\end{equation}
and
\begin{equation}
q_\mathrm{pos}(t)=0.046\left(-e^{-\frac{t}{t_\mathrm{Ni}}}+e^{-\frac{t}{t_\mathrm{Co}}}\right)\times 10^{43}\,\mathrm{erg}\,\mathrm{s}^{-1} ,
\end{equation}
 where $t_\mathrm{Ni}=8.76$\,days and $t_\mathrm{Co}=111.4$\,days. We adopt the same gamma-ray deposition function\cite{clocchiatti1997} as in the data paper\cite{rodriguez2023}, 
\begin{equation}\label{eq:fdep}
f_\mathrm{dep}(t,t_\text{esc})=1-e^{-(t_\text{esc}/t)^2}.
\end{equation}

To compute $\LTm$, the value of $t$ in equation~(\ref{eq:ET}) must be much longer than the radiation diffusion time $t_\text{diff}$, such that $E(t)\,t$ tends to zero and $L(t) \approx Q_\text{nuc}(t)$. Since at $t=t_\text{peak}$ the diffusion time is of the order of $t_\text{peak}$, while $t_\text{diff}$ drops roughly as $t^{-2}$, we can assume $E(t)\,t=0$ for $t\gtrsim 4\,t_\text{peak}$. Given that the epoch of the last luminosity measurement ($t_\text{last}$) for the SNe in our sample are at least five times longer than $t_\text{peak}$, we adopt $t=t_\text{last}$ in equation~(\ref{eq:ET}).
\ref{fig:tL_tQ} shows, beside $L(t)\,t$, the time-weighted radioactive energy deposition  $Q_\text{nuc}(t)\,t$. The figure shows that in all the SNe in our sample $Q_\text{nuc}(t)\,t \approx L(t)\,t$ long before $t_\text{last}$, implying that the exact value of $t_\text{last}$ has a negligible effect on the measurement of $\LTm$.
We calculate the integrals in $LT(t_\text{last})$ and $QT_\text{nuc}(t_\text{last})$ numerically using the trapezoidal rule.

The error on $\LTm$,  computed with error propagation, is given by
\begin{equation}\label{eq:err_logET}
\sigma_{\log \LTm} = \left[\frac{\sigma_\mu^2+\sigma_\text{BC}^2+\sigma_{\text{redd1,G}}^2+\sigma_{\text{redd1,h}}^2}{6.25}+\frac{\sigma_{LT}^2+\sigma_{QT}^2+\sigma_\text{texp1}^2}{(\LTm\,\ln{10})^2}\right]^{1/2}.
\end{equation}
Here, the term $\sigma_\text{BC}$ is the error due to uncertainties in the zero-points of the BC scales used in the data paper\cite{rodriguez2023} to compute luminosities. The mean $\sigma_\text{BC}$ values for SNe~IIb, Ib, and Ic are 0.03, 0.04, and 0.05\,mag, respectively. The terms $\sigma_{\text{redd1,G}}$ and $\sigma_{\text{redd1,h}}$ are the uncertainties due to errors in Galactic and host-galaxy reddenings, respectively, given by
\begin{equation}
\sigma_{\text{redd1},s} =  \left\lvert \langle\zeta_s\rangle+(\langle\zeta_s\rangle-\langle\zeta_s\rangle^\text{tail})\frac{\QTn(t_\text{last})}{\LTm}\right\rvert\sigma_{E(B-V)_s},
\end{equation}
with $s=\{\text{G},\text{h}\}$. Here, $\zeta_s$ is defined as in the data paper\cite{rodriguez2023}, and depends on the extinction-to-reddening ratios of the optical bands used to compute luminosities through the BC technique, and on the dependence of BC on colour. $\langle\zeta_s\rangle$ is the average of the $\zeta_s$ values between $t_\text{first}$ and $t_\text{last}$, while $\langle\zeta_s\rangle^\text{tail}$ is the mean of the $\zeta_s$ values in the radioactive tail. The typical $\langle\zeta_\text{h}\rangle$ ($\langle\zeta_\text{h}\rangle^\text{tail}$) values for SNe~IIb, Ib, and Ic are 2.3 (2.1), 2.4 (2.2), and 3.3 (3.2), respectively, while the typical $\langle\zeta_\text{G}\rangle$ ($\langle\zeta_\text{G}\rangle^\text{tail}$) is of 2.7 (2.5). The term $\sigma_{LT}$ is the uncertainty in $LT(t_\text{last})$ due to errors in luminosity, given by $\sigma_{LT}=(\sigma_{LT(t_\text{first})}^2+\sigma_{LT,I}^2)^{1/2}$, where
\begin{align}\label{eq:error_LT}
\sigma_{LT,I}=&\frac{1}{2}\,([(t_2-t_1)\,t_1\,\sigma_{L(t_1)}]^2+\sum_{i=2}^{n-1}[(t_{i+1}-t_{i-1})\,t_i\,\sigma_{L(t_i)}]^2\nonumber\\
            &+[(t_n-t_{n-1})\,t_n\,\sigma_{L(t_n)}]^2)^{1/2}
\end{align}
is the error on the integral in equation~(\ref{eq:LT_last}), $n$ being the number of data points in the luminosity light curve. The $L(t)$ errors do not include $\sigma_\mu$, $\sigma_\text{BC}$, $\sigma_{E(B-V)_\text{h}}$, and $\sigma_{E(B-V)_\text{G}}$, which are already included in equation~(\ref{eq:err_logET}). The term $\sigma_{QT}$ is the error in $QT_\text{nuc}(t_\text{last})$ due to uncertainties in $\log M_\text{Ni}$ and $t_\text{esc}$. To calculate $\sigma_{QT}$, we randomly select 10,000 pairs of $\log M_\text{Ni}$ and $t_\text{esc}$ values from their posterior probability distribution, compute $QT_\text{nuc}(t_\text{last})$ using those values, and adopt the sample standard deviation 
of the 10,000 $QT_\text{nuc}(t_\text{last})$ estimates as $\sigma_{QT}$. In the data paper\cite{rodriguez2023}, we obtained this distribution for each SN by means of the Markov Chain Monte Carlo process implemented in the package \textsc{emcee}\cite{foreman-mackey2013}, which we used to compute $\log M_\text{Ni}$ and $t_\text{esc}$. The term $\sigma_\text{texp1}$ in equation~(\ref{eq:err_logET}) is the error in $\LTm$ due to the uncertainty in $t_\text{exp}$. Changes in $t_\text{exp}$ not only affect $t$ but also the values of $\log M_\text{Ni}$ and $t_\text{esc}$ inferred from the radioactive tail of the luminosity light curves. To quantify variations in $\LTm$ due to changes in $t_\text{exp}$, for each SN we compute $\log M_\text{Ni}$ and $t_\text{esc}$ in the same manner as in the data paper\cite{rodriguez2023} but using $t_\text{non-det}$ and $t_\text{detect}$ to estimate the explosion time. These epochs correspond to the lowest and greatest possible value for $t_\text{exp}$, respectively. Then, we compute $\LTm$ using $t_\text{exp}+\Delta t_\text{exp}$ as the explosion time, where $\Delta t_\text{exp}$ is equal to $t_\text{non-det}-t_\text{exp}$ and $t_\text{detect}-t_\text{exp}$. \ref{fig:dLT-nuc_dt}a shows the relative change in $\LTm$ ($\Delta \LTm/\LTm$) against $\Delta t_\text{exp}$. The distribution, which is well represented by a straight line with intercept 0.002 and slope $-0.067$\,d$^{-1}$,
indicates that $\LTm$ decreases as the time of explosion advances. Based on 
this, we adopt $\sigma_\text{texp1}=0.067\, \lvert\LTm\rvert \,\sigma_{t_\text{exp}}\,\text{d}^{-1}$.

The $\LTm$ values and their errors are listed in \ref{table:LT-nuc}. The typical $\log \LTm$ uncertainty is of 0.18\,dex, corresponding to a relative $\LTm$ error of 41\%. 
On average, 71\%, 12\%, and 9\% of the $\LTm$ error is induced by uncertainties in $E(B-V)_\text{h}$, $\mu$, and $\QTn(t_\text{last})$, respectively. 

\subsection*{Estimating $\LTm/LT_{100}$ from the observations}
Given that host galaxy reddening and distance dominate the $\LTm$ error budget, it is useful to compute $\LTm/LT_{100}$, where $LT_{100}\equiv LT(t=100\,\text{d})$. This ratio, which corresponds to the contribution of non-radioactive energy sources, $\LTm$, to the time-weighted luminosity integrated over the first 100~days after explosion, is independent of the distance and of the zero-point of the BC scale, and less dependent on reddening than $\LTm$. For 23~SNe in our sample with $t_\text{last}<100\,\text{d}$, we compute $LT_{100}$ using $LT_{100}=LT(t_\text{last})+\int_{t_\text{last}}^{100\,\text{d}}Q_\text{nuc}(t)\,t\,dt$, assuming a conservative relative error of 20\% for the integral.

The error in $\LTm/LT_{100}$, computed with error propagation, is given by
\begin{align}
\sigma_{\LTm/LT_{100}} = &\frac{\LTm}{LT_{100}}\left[\sigma_\text{redd2,G}^2+\sigma_\text{redd2,h}^2+\frac{(\sigma_{LT}^2+\sigma_{QT}^2)}{\LTm^2}+\frac{\sigma_{LT_{100}}^2}{LT_{100}^2}\right.\nonumber\\
&\left.+\frac{\sigma_\text{texp2}^2}{(\LTm/LT_{100})^2}\right]^{1/2},
\end{align}
where
\begin{equation}
\sigma_{\text{redd2},s} =\frac{\ln 10}{2.5}\left\lvert(\langle\zeta_s\rangle-\langle\zeta_s\rangle^\text{tail})\frac{\QTn(t_\text{last})}{\LTm}\right\lvert\sigma_{E(B-V)_s},
\end{equation}
$\sigma_{LT_{100}}$, similar to $\sigma_{LT}$, is the uncertainty in $LT_{100}$ due to errors in luminosity, and $\sigma_\text{texp2}$ is the error in $\LTm/LT_{100}$ due to uncertainties in $t_\text{exp}$. To quantify variations in $\LTm/LT_{100}$ due to changes in $t_\text{exp}$, we proceed in the same manner as we did in the case of $\LTm$. \ref{fig:dLT-nuc_dt}b shows changes in $\LTm/LT_{100}$ ($\Delta \LTm/LT_{100}$) normalised to $\LTm/LT_{100}$ against $\Delta t_\text{exp}$. The observed distribution,
which is well represented by a straight line with intercept $-0.003$ and slope $-0.047$\,d$^{-1}$,
indicates that $\LTm/LT_{100}$ decreases as the explosion epoch advances. 
Based on this,
we adopt $\sigma_\text{texp2}=0.047\,\lvert\LTm/LT_{100}\rvert\,\sigma_{t_\text{exp}}\,\text{d}^{-1}$.

The $\LTm/LT_{100}$ estimates and their errors are listed in \ref{table:LT-nuc}. The mean $\LTm/LT_{100}$ uncertainty is of 0.028, corresponding to a relative $\LTm/LT_{100}$ error of 23\%. On average, 42\%, 31\%, 13\%, and 13\% of the $\LTm/LT_{100}$ error is induced by uncertainties in $\QTn(t_\text{last})$, $E(B-V)_\text{h}$, $LT(t_\text{last})$, and $t_\text{exp}$ respectively.

\subsection*{Correlations}
\begin{figure*}
\includegraphics[width=1.0\textwidth]{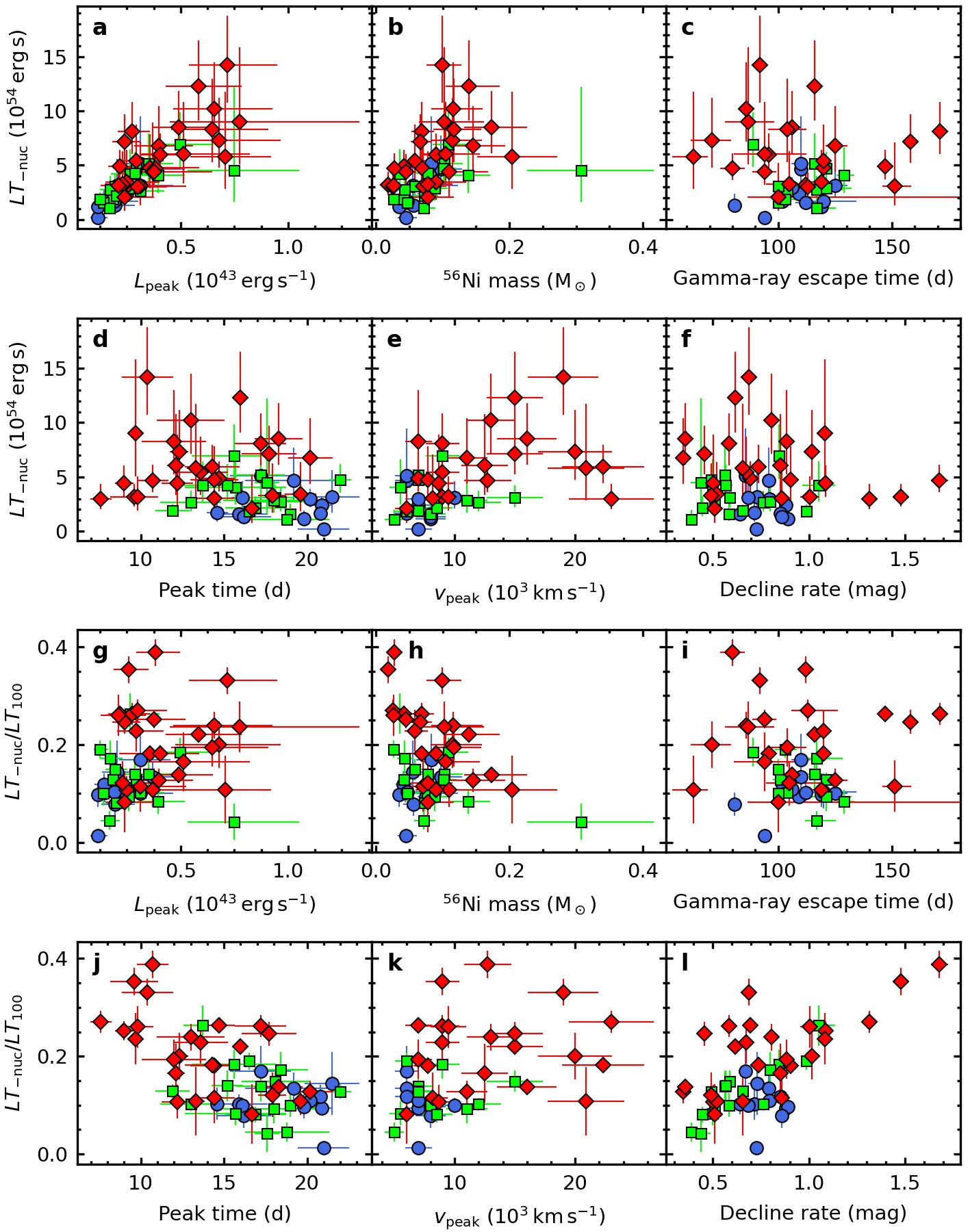}
\caption{{\bf Correlations between the non-radioactive contribution and various quantities.} $\LTm$ (\textbf{a--f}) and $\LTm/LT_{100}$ (\textbf{g--l}) against peak luminosity (\textbf{a, g}), $^{56}$Ni mass (\textbf{b, h}), gamma-ray escape time (\textbf{c, i}), peak time (\textbf{d, j}), ejecta velocity at peak time (\textbf{e, k}), and decline rate (\textbf{f, l}). Error bars here and in subsequent figures denote $1\,\sigma$ errors.}
\label{fig:LT-nuc_vs_x}
\end{figure*}

To investigate our results and to test for systematic errors that may artificially drive them, we have searched for correlations among the observables. \ref{fig:LT-nuc_vs_x} shows $\LTm$ and $\LTm/LT_{100}$ plotted against $L_\text{peak}$, $M_\text{Ni}$, $t_\text{esc}$, $t_\text{peak}$, $v_\text{peak}$, and $\Delta m_{15}$. For each of these pairs of variables, we compute Pearson's correlation coefficients ($r$) and the corresponding $P$ values.
We focus on strong ($\lvert r\rvert\geq0.7$) and moderate ($0.5\leq \lvert r\rvert<0.7$) correlations with $P$ values $\leq0.05$, which are statistically significant at the $>95$\% level. For SNe~IIb, Ib, Ic, and SE~SNe as a whole, we find a strong correlation between $\LTm$ and $L_\text{peak}$
($r$ values of 0.89, 0.70, 0.73, 0.79, respectively, and $P\leq0.003$).
For $\LTm$ versus $M_\text{Ni}$, we find a strong correlation for SNe~IIb ($r=0.82$, $P=0.001$) and a moderate correlation for SNe~Ic ($r=0.50$, $P=0.01$). 
The previously suggested\cite{sharon2020}  possible correlation between $\LTm$ and $t_\text{esc}$ is not reproduced in our analysis ($r=-0.06$, $P=0.70$), and is likely a consequence of the small number (nine) of SE~SNe used in that previous work. Indeed, using the $\LTm$ and $t_\text{esc}$ values of the eight SNe in common between the present work and the previous sample\cite{sharon2020}, we obtain $r=0.78$ ($P=0.02$), corresponding to a strong correlation.
For SNe~Ib we find a strong correlation between $\LTm/LT_{100}$ and $\Delta m_{15}$ ($r=0.88$, $P<0.001$) and a moderate correlation between $\LTm/LT_{100}$ and $M_\text{Ni}$ ($r=-0.53$, $P=0.03$), while for SNe~Ic we find
moderate correlations between $\LTm/LT_{100}$ and $M_\text{Ni}$, $t_\text{peak}$, and $\Delta m_{15}$ ($r$ values of $-0.57$, $-0.58$, and 0.69, respectively, and $P\leq0.002$). For SE~SNe as a whole, we find moderate correlations between $\LTm/LT_{100}$ and $t_\text{peak}$ ($r=-0.63$, $P<0.001$) and between $\LTm/LT_{100}$ and $\Delta m_{15}$ ($r=0.63$, $P<0.001$). Given that $\LTm$ 
correlates with $L_\text{peak}$
and that $LT_{100}=QT_\text{nuc}(t=100\,\text{d})+\LTm$, 
where $QT_\text{nuc}(t=100\,\text{d})$ is proportional to $M_\text{Ni}$, the quantity $\LTm/LT_{100}$ is roughly proportional to $L_\text{peak}/M_\text{Ni}$. Therefore, the correlation between $\LTm/LT_{100}$ and $t_\text{peak}$ arises from the correlation between $L_\text{peak}/M_\text{Ni}$ and $t_\text{peak}$ (known as the peak time-luminosity relation) that has been observed for SE~SNe (e.g. see the data paper\cite{rodriguez2023}), while the correlation between $\LTm/LT_{100}$ and $\Delta m_{15}$ arises because $L_\text{peak}/M_\text{Ni}$  correlates not only with $t_\text{peak}$ but also with $\Delta m_{15}$\cite{rodriguez2023}.

\subsection*{Selection bias}
The selection criteria we have used to construct our SN sample are mainly connected to the temporal coverage of each SN light curve. This process could, in principle, somehow create a sample that is biased in its properties compared to the parent population of all SE~SNe. To roughly evaluate the strength of any such bias, we compare our sample with the volume-limited (VL) samples used in the data paper\cite{rodriguez2023}. These samples consist of 18~SNe~IIb, 19~SNe~Ib, and 29~SNe~Ic with $\mu\leq32.8$, 33.3, and 34.0\,mag, respectively, where the upper limits were selected such that the selection bias affecting the VL samples is relevant only for a small fraction of SNe with low luminosity. For the comparison, we use $M_{r,\text{peak}}$.

\begin{figure*}
\includegraphics[width=1.0\columnwidth]{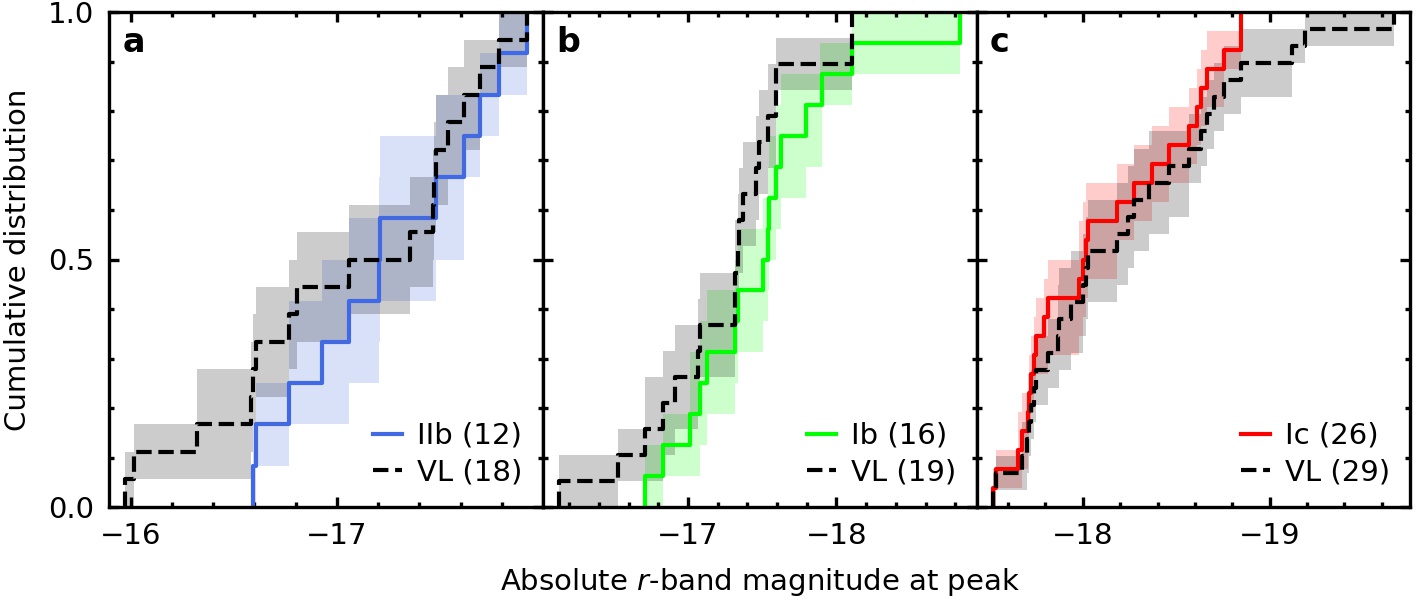}
\caption{{\bf Peak luminosity comparison of our sample to a volume-limited sample.} Cumulative distributions for the absolute $r$-band magnitudes at peak of the SNe~IIb (\textbf{a}), Ib (\textbf{b}), and Ic (\textbf{c}) in our sample (solid lines) and in the volume-limited samples of ref.\cite{rodriguez2023} (dashed lines). Shaded regions represent 68\% confidence intervals computed by bootstrap resampling (10,000 samples). Numbers in parentheses are the sample sizes.}
\label{fig:Mrpeak_cdf}
\end{figure*}

\ref{fig:Mrpeak_cdf} shows the cumulative distributions for the $M_{r,\text{peak}}$ values of the SNe~IIb, Ib, and Ic in our sample and in the corresponding VL samples. To test, for each SN subtype, whether the $M_{r,\text{peak}}$ distribution for our sample and that for the VL sample are drawn from a common unspecified distribution (the null hypothesis), we use the two-sample Anderson-Darling test\citep{scholz1987}. For 
each SN subtype we obtain Anderson-Darling $P$ values
greater than 0.38 (i.e., over the 0.05 critical level), meaning that the null hypothesis cannot be rejected. Assuming that the VL samples of the data paper\cite{rodriguez2023} are adequate approximations for complete samples, then any selection bias affecting our SN sample is also not significant.

\subsection*{Robustness of the results}
Our main result is that $\LTm$ is significantly greater than zero and a non-negligible fraction of $LT_{100}$. As this result has far-reaching implications, it is important to consider possible unaccounted systematic errors that are responsible for it. To artificially obtain our result, we would need to systematically overestimate $LT$ by 10--20\% (depending on the SN type), and/or underestimate $M_\text{Ni}$ by a similar amount. The fact that $\LTm/LT_{100}$ is non-negligible for all SNe implies that our result is independent of the adopted distances and BC zero-points. Moreover, the systematic error should affect all the SNe in our sample, and be correlated with the SN type. We note also that, in a previous study\cite{sharon2020}, the authors found that $\LTm$ is consistent with zero for SNe~Ia, and therefore if there is a systematic error in our method it needs be restricted to SE~SNe. Below, we examine a number of possible errors, but find no error that is likely to explain our results.  \\

\begin{figure*}
\includegraphics[width=1.0\columnwidth]{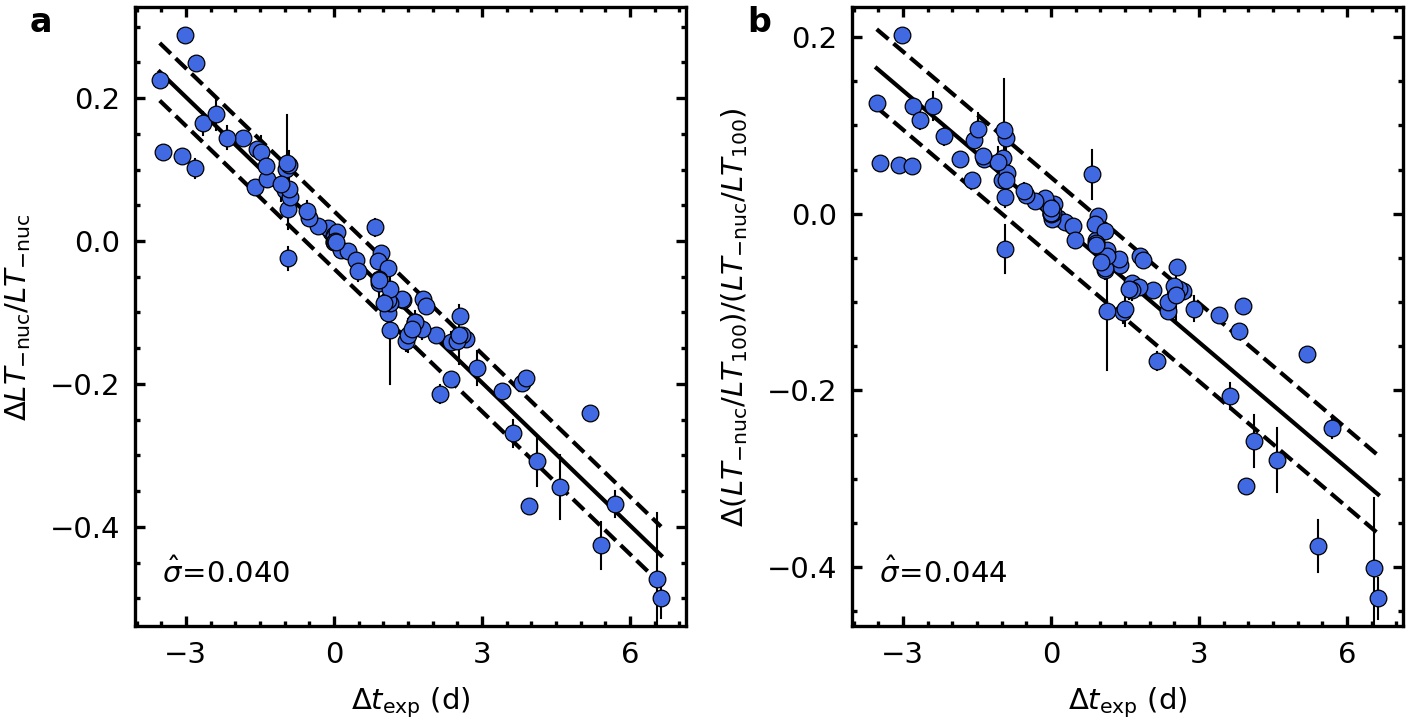}
\caption{{\bf The effect of change in the estimated explosion time.} Relative change in $\LTm$ (\textbf{a}) and $\LTm/LT_{100}$ (\textbf{b}) against change in explosion time. Negative (positive) $\Delta t_\text{exp}$ values are quantities computed using $t_\text{non-det}$ ($t_\text{detect}$) as explosion time. Error bars are $1\,\sigma$. Solid lines are straight line fits and dashed lines are $\pm1\,\hat\sigma$ limits, where $\hat\sigma$ is the sample standard deviation.}
\label{fig:dLT-nuc_dt}
\end{figure*}

\noindent{\it Explosion time:}\\
We first consider the effect of a systematic error in the estimated explosion time. Based on \ref{fig:dLT-nuc_dt} we find that later values of $t_\text{exp}$ reduce $\LTm$ and $\LTm/LT_{100}$ while earlier values increase them. Thus, our $\LTm$ and $\LTm/LT_{100}$ estimates could be overestimated if the explosion epochs that we use are systematically too early. In the extreme case where $t_\text{exp}=t_\text{detect}$ (the latest possible value for $t_\text{exp}$), the mean $\LTm/LT_{100}$ values for SNe~IIb, Ib, and Ic reduce to $0.10\pm0.01$, $0.11\pm0.01$, and $0.19\pm0.01$ ($1\,\sigma$ errors), respectively, which are only slightly lower than our original results and are still  non-zero at high significance. Thus, our finding that SE~SNe have non-zero $\LTm$ values is not a result of the adopted explosion epochs.\\

\noindent{\it Bolometric correction:}\\
An additional source of error can be the bolometric correction (BC). Here, there are two possibilities. The first is that we have overestimated the luminosity around the peak, which leads to an overestimate of $LT$. We can check this by comparing our adopted luminosity to the integrated light in SNe with observations in all bands from the UV (based on {\it Neil Gehrels Swift Observatory} UVOT data) to the near-IR $K$ band, which gives a lower limit on the luminosity. There are 10~SNe in our sample with such data, and in each of them the observed integrated light is $>95\%$ of the peak luminosity that we have adopted. Thus, an overestimated BC near the peak is unlikely to explain the values of $\LTm$.

A second option is an underestimate of the BC during the radioactive tail phase, that leads to an underestimate of $M_\text{Ni}$. For three SNe with {\it Swift} data during the tail we find that the contribution of the UV to the bolometric flux is less than 2\%. Therefore, it is unlikely that there is a missing relevant contribution in the UV during the tail.
To explore emission in the mid-infrared, we use SNe with {\it Spitzer} observations. There are only a handful of SE~SNe with such observations during the first $\sim 100$~days, with the best studied case being SN~2011dh\citep{tinyanont2016}. These SNe display mid-IR emission that, during the first 100~days after the peak, decreases with time more slowly than the emission in other bands. This slow drop has been attributed to emission by warm dust, and its fraction as part of the total luminosity grows slowly with time\citep{szalai2019}. The heating source of this dust is probably radiative, either by the shock breakout or from the main SN emission, most likely near the peak. The time delay between this heating and the observed dust emission could be the result of light travel time. In any case, the light seen in the dust component at any given time of the light curve is a robust upper limit, and most likely results in a significant overestimate of the instantaneous emission that is lost to dust heating, and is therefore missing from our estimate of the bolometric luminosity. Mid-IR measurements are available for two of the SNe in our sample, 2011dh (IIb) and 2014L (Ic). A comparison of the dust luminosity (as estimated with a blackbody fit\cite{tinyanont2016}) to the bolometric luminosity, for SN~2011dh, shows that the dust luminosity is about 1\% of the total luminosity at day 17, and it is growing with time, reaching slightly less that 5\% at day 84. A similar comparison for SN~2014L shows that the dust luminosity grows from 2.6\% to 3.5\% from day 40 to 67.

Our conclusion is that while dust absorption may have some effect on the estimated $^{56}$Ni mass, this effect is small. It cannot be larger than a few per~cent and most likely it is about 1\% or less (at least in SN~2011dh and SN~2014L). This is certainly true if the dust optical depth is constant with time, and it is also true if the optical depth grows with time (e.g., due to dust formation), since a growing dust opacity mimics thermalization losses and therefore it is compensated, at least partially, by an underestimate of $t_\text{esc}$. Thus, it seems highly unlikely that systematic errors in the BC are the source of the positive values of $\LTm$ that we find.\\

\noindent{\it Host galaxy reddening:}\\
The spectral energy distributions of SE~SNe are bluer near maximum light than at later epochs (e.g. ref.\cite{stritzinger2018}), so reddening corrections in the radioactive-tail phase will be lower than at epochs near peak. Overcorrecting for reddening can therefore produce a greater increment in $LT$ than in $\QTn$, leading to an overestimated, and perhaps systematically non-zero, $\LTm$.

\begin{figure*}
\includegraphics[width=1.0\columnwidth]{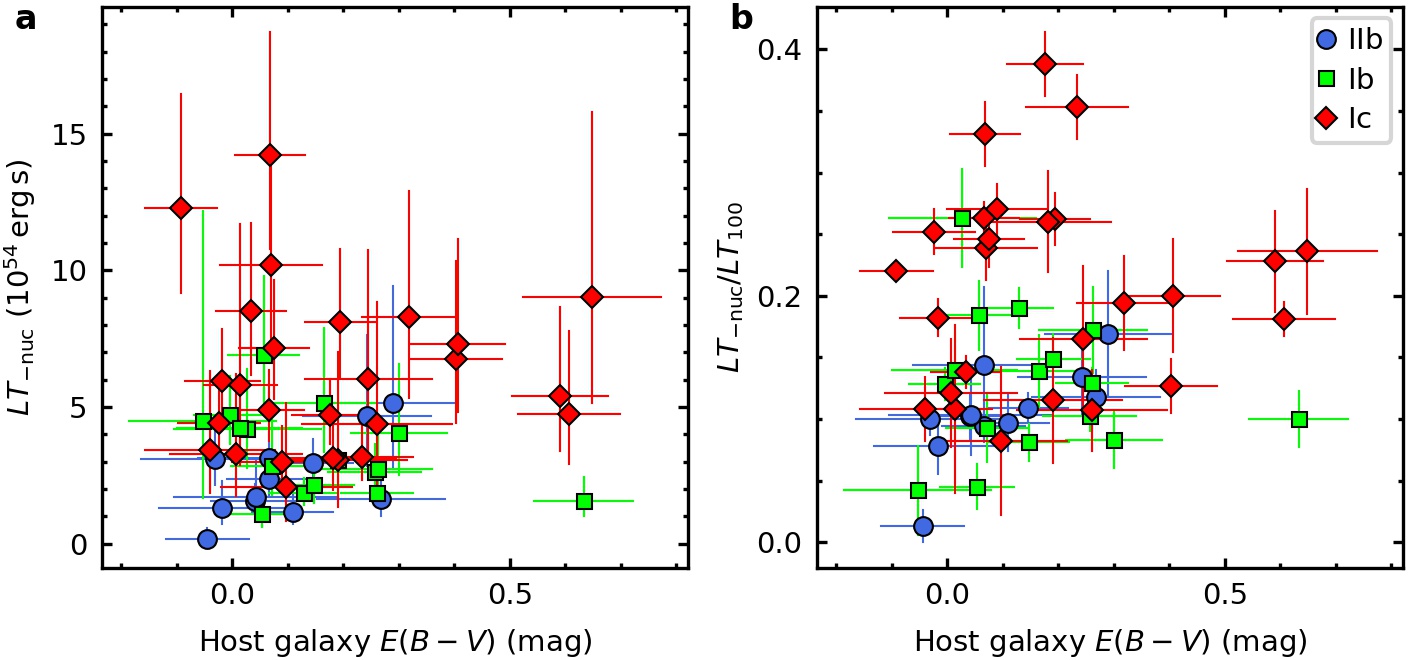}
\caption{{\bf The effect of extinction.} $\LTm$ (\textbf{a}) and $\LTm/LT_{100}$ (\textbf{b}) against host galaxy reddening.}
\label{fig:LTm_LTmLT100_vs_EhBV}
\end{figure*}

\ref{fig:LTm_LTmLT100_vs_EhBV} shows $\LTm$ and $\LTm/LT_{100}$ against $E(B-V)_\text{h}$. We see that SNe with high values of $\LTm$ and $\LTm/LT_{100}$ do not necessarily have high $E(B-V)_\text{h}$ values. In addition, using only SNe with low host-galaxy reddening ($E(B-V)_\text{h}<0.1$\,mag), we obtain mean $\LTm/LT_{100}$ values of $0.09\pm0.01$ (7~SNe~IIb), $0.13\pm0.03$ (7~SNe~Ib), and $0.20\pm0.02$ (13~SNe~Ic), values similar to those for the full sample, and still greater than zero at a high significance. Our non-zero $\LTm$ estimates are therefore not due to a systematic overestimate of the adopted host-galaxy reddening values. \\

\noindent{\it Deposition function:}\\
In this work we have adopted the deposition function\cite{clocchiatti1997} in equation~(\ref{eq:fdep}) to describe the fraction of the generated gamma-ray energy deposited in the ejecta. Recently,  a more versatile deposition function has been proposed\cite{sharon2020}, given by $f_\text{dep}^\text{S\&K}=(1+(t/t_\text{esc})^n)^{-2/n}$, where $n$ is an additional free parameter that controls the sharpness of the transition between the gamma-ray optically thick and thin regimes. The $^{56}$Ni masses calculated with this deposition function are, on average, 14\% larger than those computed assuming equation~(\ref{eq:fdep})\citep{rodriguez2023}. Since larger $^{56}$Ni masses produce larger $\QTn$ values, our $\LTm$ and $\LTm/LT_{100}$ estimates could be overestimated if $f_\text{dep}^\text{S\&K}$ is the true deposition function for SE~SNe.

\begin{figure*}
\includegraphics[width=1.0\columnwidth]{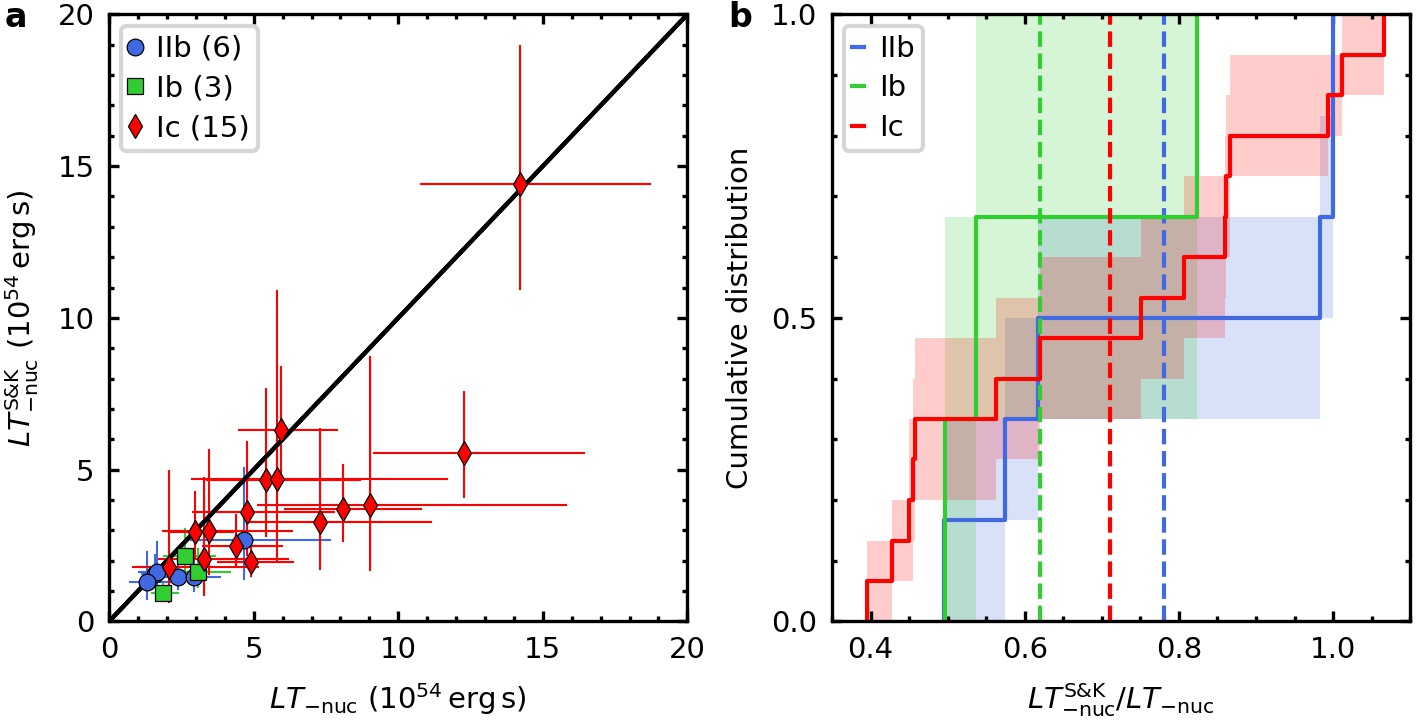}
\caption{ {\bf The effect of the radioactive energy deposition function.} \textbf{a}. $\LTm$ computed with the deposition function of Sharon \& Kushnir against the $\LTm$ estimates reported in this work. The solid line is a one-to-one correspondence. \textbf{b}. Cumulative distribution for the ratio of $\LTm$ computed with the deposition function of Sharon \& Kushnir to $\LTm$ reported in this work. Shaded regions represent 68\% confidence intervals computed by bootstrap resampling (10,000 samples). Vertical dashed blue, green, and red lines indicate mean values for SNe~IIb, Ib, and Ic, respectively}
\label{fig:LTm_comparison}
\end{figure*}

\ref{fig:LTm_comparison}a shows $\LTm$ calculated assuming $f_\text{dep}^\text{S\&K}$ as the deposition function ($\LTm^\text{S\&K}$) against the $\LTm$ estimates reported in this work. For the 24~SE~SNe with reasonable $n$ values ($n>1$; see ref.\cite{sharon2020}) shown in the figure, we see that the $\LTm^\text{S\&K}$ values are mostly lower than our $\LTm$ estimates but still greater than zero. \ref{fig:LTm_comparison}b shows the cumulative distribution for the ratio of $\LTm^\text{S\&K}$ to our $\LTm$ estimates. The mean $\LTm^\text{S\&K}/\LTm$ values are $0.78\pm0.10$ (6~SNe~IIb), $0.62\pm0.10$ (3~SNe~Ib), and $0.71\pm0.06$ (15~SNe~Ic), which are statistically consistent with each other. Therefore, if $f_\text{dep}^\text{S\&K}$ is the true deposition function for SE~SNe, then our $\LTm$ and $\LTm/LT_{100}$ estimates are overestimated, on average, by a factor of 1.4. In particular, the mean $\LTm/LT_{100}$ values reduces to $0.09\pm0.01$, $0.08\pm0.01$, and $0.15\pm0.02$ for SNe~IIb, Ib, anc Ic, respectively, which are still significantly greater than zero. Thus, it seems unlikely that our non-zero $\LTm$ estimates result from our adopted deposition function. Yet, given the sensitivity of the exact value of $\LTm$ and $\LTm/LT_{100}$ to the adopted deposition function, in the unlikely case that the deposition in reality is very different than the forms suggested by theory, it may be the source of the positive values of $\LTm$.  \\

\noindent{\it Asphericity:}\\
Equation~(\ref{eq:ET}) is independent of the geometry of the ejecta. However, our estimates of $L$ and $M_{\rm Ni}$ from the observations do assume spherical symmetry, and a deviation from spherical symmetry could lead to an over- or underestimate of both $L$ and $M_{\rm Ni}$. The errors in the estimates of $L$ and $M_{\rm Ni}$ are not necessarily the same and this may introduce a non-zero value of $\LTm$. However, as explained below, for some SNe asphericity could potentially increase the value of $\LTm$, but then for a similar number of SNe it would then reduce its value. The fact that all SNe show a positive value of $\LTm$ implies that asphericity is not a dominant systematic effect.

If the SN emission is not isotropic then, since we assume that $L$ is proportional to the observed flux, some observers will infer a peak luminosity that is higher than the actual one, while others will infer a luminosity that is lower. The same will happen during the radioactive tail phase, with observers who measure higher values of $L$ during peak, also measuring a higher value of $M_{\rm Ni}$. However, since the optical depth of the entire ejecta drops with time, the emission becomes more isotropic. Thus, the over- and underestimates of $M_{\rm Ni}$ are closer to the true (direction-averaged) values than the over- and underestimate of $L$, and thus of $LT$, near peak time. The result is that even if radioactive decay were the only energy source, an observer who overestimates $LT$ by some factor is expected to overestimate $M_{\rm Ni}$ by a lower factor and thus to measure a positive $\LTm$. An observer of the same SN from a different viewing angle, who underestimates $LT$ by some factor will underestimate $M_{\rm Ni}$ by a lower factor and hence will measure a negative $\LTm$. As in our sample there is not a single SN with a negative value of $\LTm$, it is unlikely that that the entire effect we measure results from asphericity, when in fact $\LTm$ is zero.


\subsection*{Interaction with a circumstellar medium}
CSM interaction is an efficient source of energy for SN emission. The emission from CSM interaction depends strongly on the optical depth of the medium.  If the medium is optically thin for absorption then the emission is nonthermal, extending from radio to X-rays. If it is optically thick to absorption, then it can peak in the UV/optical/IR. However, in that case, during the time that the interaction takes place the transfer of the radiation through the unshocked CSM generates narrow emission lines, which are the hallmark of CSM interaction. 

Observations of normal SE SNe, such as  those that constitute our sample, strongly disfavor the possibility that interaction contributes a significant fraction of the  UV/optical/IR emission seen in most or all such SNe. This is supported by three independent lines of evidence:\\
a) Many normal SE SNe, including some of the SNe in our sample, are accompanied by synchrotron radio emission that arises from CSM interaction\cite{chevalier1998}. This emission arises from interaction of the fastest moving ejecta, at velocities of 30,000\,km\,s$^{-1}$ and more, with an optically thin wind. The contribution of this emission to the SN UV/optical/IR light is completely negligible. There are no signs in radio or X-ray for an additional interaction that could contribute to the UV/optical/IR emission.  \\
b) As explained above, a significant contribution of CSM interaction to the UV/optical/IR light, without displaying bright emission in the radio and/or X-rays, requires interaction with an optically thick medium. However, in that case narrow lines are expected, but which are observed in none of the SNe in our sample.\\
c) During CSM interaction, the radiation is generated at the leading edge of the ejecta (where the interaction takes place) and the arrival time of photons to the observer is determined by the diffusion time through the unshocked CSM. There is no reason to expect that the time of the peak of this emission coincides with the peak of the radioactive emission, which is determined by the diffusion time through the entire ejecta. The fact that in all SNe in our sample there is a single peak in the light curve, which coincides with the expected contribution of radioactive power, argues that CSM interaction is not a dominant source of light, at least in the majority of these SNe. Note that the above considerations are distinct from those related to the time at which engine-powered emission is released. The reason is that, unlike CSM interaction, the engine deposits its energy at the base of the ejecta and (assuming energy deposition before the peak) this energy is released together with the peak of the radioactive powered emission, i.e. when the diffusion time trough the ejecta is comparable to the expansion time.   
Our conclusion is that CSM interaction cannot be the source of $\LTm$, at least in most (and most likely all) of the SNe in our sample.

\subsection*{Progenitor radius assuming that $ET \approx \LTm$}
If the main energy source of $\LTm$ is cooling emission, then we can estimate the progenitor radius (or more specifically an integral over its pre-explosion mass density  distribution\cite{shussman2016}). The value of $ET$ is determined by the progenitor properties and the explosion energy. This has been discussed in detail\cite{shussman2016}, and it was shown that $ET =A (E_{\rm exp} M_{\rm ej})^{1/2}R_\text{prog}$, where $E_{\rm exp}$ is the total explosion energy, $M_{\rm ej}$ is the ejecta mass, $R_\text{prog}$ is the progenitor radius, and $A$ is a proportionality coefficient that depends on the exact density distribution of the pre-explosion progenitor and on deviations from spherical symmetry. This relation can be understood as follows. Each fluid element reaches its homologous expansion phase roughly when it arrives at double the radius at which it was shocked. Its internal energy at this time is proportional to its kinetic energy (where the latter  dominates the former by up to an order of magnitude). For the bulk of the progenitor mass, the internal energy is proportional to $E_{\rm exp}$,  and the time at which it becomes homologous is roughly $R_\text{prog}/(E_{\rm exp}/M_{\rm ej})^{1/2}$, so $ET \equiv E(t_0)t_0 \propto E_{\rm exp} R_\text{prog}/(E_{\rm exp}/M_\text{ej})^{1/2}=(E_\text{exp} M_{\rm ej})^{1/2}R_\text{prog}$.

The value of $A$ reflects the distribution of mass within the progenitor and the deviations of the explosion from spherical symmetry, if there are any. Progenitors with mass that is concentrated toward the core will have a smaller $A$ values, compared to a progenitor with mass that is spread out closer to the surface. It was found that for a large range of red supergiant progenitors calculated with the stellar-evolution code \textsc{mesa}\citep{paxton2011}, the average $A$ value is $0.15$\cite{shussman2016} . Carrying out a similar calculation for analytic profiles of stars with radiative envelopes (as expected for SE~SNe progenitors), where the mass distribution is slightly more concentrated than in stars with convective envelopes, we find $A \approx 0.1$. We can therefore estimate the radius of the progenitor with the following relation:
\begin{align}\label{eq:WR_ET}
    \LTm & \approx  ~ET ~ \approx~ 0.1  (E_{\rm exp} M_{\rm ej})^{1/2} R_\text{prog} \nonumber\\
    &= 3 \times 10^{52}\,{\rm erg\,s}\left(\frac{E_{\rm exp}}{10^{51}\,{\rm erg}}\frac{M_\text{ej}}{5\,M_\odot}\right)^\frac{1}{2}  
    \frac{R_\text{prog}}{10^{11}\,{\rm cm}}.
\end{align}
Note that $R_\text{prog}$ is not necessarily the photospheric radius of the star, but rather the radius where most of the mass is concentrated (see discussion below).

In order to evaluate $R_\text{prog}$ in the case that $ET=\LTm$, we use a rough estimate of $M_\text{ej}$ and $E_\text{exp}$ from the observables: $M_\text{ej} \approx (\beta/2) \kappa^{-1} c\,t_{\rm peak}^2 v_{\rm peak}$ and $E_\text{exp} \approx 0.3\,M_\text{ej}\,v_\text{peak}^2$, where $\beta$ is an integration coefficient that depends (among other things) on the ejecta density and velocity distributions, $\kappa$ is an effective opacity and $c$ is the speed of light. Here we use, following precedent\cite{prentice2019}, $\beta=13.7$ and $\kappa=0.07\,{\rm cm^2\,g^{-1}}$. Plugging these relation into equation~(\ref{eq:WR_ET}) we obtain
\begin{align}\label{eq:Rprog}
    R_\text{prog} & \sim \frac{\LTm}{5c\,t_{\rm peak}^2\,v_{\rm peak}^2}\nonumber\\
    &\approx  10^{13}\,{\rm cm}\,\frac{\LTm}{3 \times 10^{54}\,{\rm erg\,s}}
        \left(\frac{t_{\rm peak}}{15\,{\rm d}}\,\frac{v_{\rm peak}}{10^4\,{\rm km \,s^{-1}}}\right)^{-2} .
\end{align}
Given that the approximations of the explosion energy and the ejecta mass are crude, we assess the accuracy of this relation at about an order of magnitude. Applying equation~(\ref{eq:Rprog}) to the SNe in our sample, we find that, if the non-radioactive energy source is dominated by cooling emission, then the progenitors have radii in the range 
\begin{equation}
    R_{\rm prog} \sim 5 \times 10^{12} - 5\times 10^{13}\,\text{cm}.
\end{equation}
We re-emphasize that $R_{\rm prog}$ is the radius where most of the ejecta resides before the explosion, rather than the photospheric radius. Thus, in progenitors that have a low-mass extended envelope, $R_\text{prog}$ is the core radius, which can be  smaller than the photospheric radius by orders of magnitude. As discussed below, this distinction is especially important in the context of SE~SNe, since some of their progenitors, especially of type IIb, have such a low-mass extended envelope. 

Stellar evolution models of SE~SNe progenitors\cite{yoon2010,laplace2020} predict the following structures. If the star retains a non-negligible hydrogen envelope ($\sim 0.01-0.1\,M_\odot$), then this envelope expands to hundreds of solar radii, while the helium core retains a radius of $\sim 10^{11}$\,cm. Such progenitors are expected to produce Type IIb SNe. If hydrogen is removed below some mass threshold then helium is abundant in the envelope and the envelope radius depends on its core mass. For a core mass larger than about $3\,M_\odot$, the envelope is compact and the photospheric radius is at most a few solar radii. For core mass lower than about $3\,M_\odot$ the envelope expands to $\sim 100\,R_\odot$ and the progenitor becomes a helium giant. However, similar to the case above of progenitors with a low-mass hydrogen envelope, here too the extended envelope contains only a small fraction of the mass while over 95\% of the stellar mass is confined to no more than a few solar radii\cite{yoon2010}. Thus, since $R_\text{prog}$, which dominates the contribution to $ET$, reflects the radius of the core, we conclude that the constraints found in equation~(\ref{eq:Rprog}) are inconsistent with stellar-evolution models of SE~SN progenitors.

Observationally, in almost all cases of SE~SNe in which progenitor candidate stars have been identified in pre-explosion images, the SN was of type IIb. These are SNe that have retained a small amount of hydrogen, and therefore the expectation is that the progenitor has a massive core, surrounded by a low-mass envelope. This expectation is supported by the observations. The observational signature of such a progenitor is a pre-explosion photospheric radius of a supergiant, and a double-peaked SN light curve, where the first peak is due to shock cooling and the second is dominated by $\Ni$\cite{nakar2014}. Well known examples are the Type~IIb~SNe ~1993J and 2011dh. Pre-explosion images found that they had supergiant progenitors with photospheric radii of about $10^{13}$\,cm\cite{Aldering1994A,vandyk2013}. The cooling emission of the envelope in these SNe peaked within a few days of the explosion, indicating that the measured radius of each progenitor represents the photosphere of a low-mass hydrogen envelope ($\lesssim 0.1\,M_\odot$), while most of the ejecta mass originated from a much more compact core\cite{Blinnikov1998,bersten2012}, even if the radius of this core is not tightly bound by observations. In the single unambiguous case of a pre-explosion detection of a Type~Ib SN progenitor, the estimated pre-explosion radius is $\sim 3 \times 10^{12}$\,cm, and here too the theoretical models suggest that the progenitor had a massive core with a low-mass extended envelope\citep{gilkis2022}. We conclude that the observations support the stellar evolution models, according to which the value we measure for $\LTm$ cannot be dominated by cooling envelope emission. Nonetheless, as the observations do not provide direct constraints on the core radii of the progenitors, we cannot completely rule out this possibility.


Fortunately, the bolometric light curves of the SNe themselves allow us to carry out a direct observational consistency test of the hypothesis that $\LTm=ET$. If there were no $\Ni$ in the ejecta (and no engine) then all of the emission would be cooling emission. In such a case, the luminosity would be very high during shock breakout ($\sim 10^{44}$\,erg\,s$^{-1}$). Then, during the ensuing cooling emission phase, after a short episode of a rapid decay, the luminosity would start decaying slowly, roughly as $t^{-0.35}$\citep{nakar2010}. This phase of shallow decay stops once the photosphere's location is affected by recombination, at which point the luminosity evolution flattens even further (or even rises very slowly), roughly to a plateau with a luminosity $L_{\rm plat}$. The plateau ends at time $t_{\rm plat}$ with an abrupt drop, as the diffusion time through the entire ejecta becomes comparable to the dynamical time (i.e. the time since explosion). This behaviour of the predicted light curve of a $\Ni$-free SE~SN can be seen in numerical models\cite{dessart2011}. When observed, it implies that, with no $\Ni$ and no engine, $ET=\LTm = LT \approx L_{\rm plat}t_{\rm plat}^2/2$. 
When the ejecta does contain $\Ni$, the early light curve remains unchanged, as the cooling emission dominates over the radioactive power.  The $\Ni$ power starts affecting the  emission only about one day, or even a few days, after the explosion, the exact time depending on the degree of $\Ni$ mixing in the external layers of the ejecta, and on the progenitor radius. Thus, as long as the cooling emission dominates the luminosity, the light curve decays  or remains roughly constant with time. Once the $\Ni$ power begins to dominate, the luminosity increases with time up to the main peak a few weeks past explosion. This behaviour of the light curve (including a direct comparison between SE~SNe light curve with and without $\Ni$) can be seen in numerical simulations\cite{dessart2011}, and a detailed discussion\cite{piro2013} of the early light curve from $\Ni$-powered SNe exists as well. From that discussion we see that, at any time before the peak of a SN that contains $\Ni$, the luminosity $L$ is  larger than, or at most comparable to, the cooling emission of a similar SN that does not contain any $\Ni$, i.e. $L \gtrsim L_{\rm plat}$ for any $t<t_p$. An additional effect of the $\Ni$ power is that it raises the temperature of the ejecta, compared to the $\Ni$-free case, and therefore recombination is delayed. This results in higher optical depth of the ejecta and therefore $t_p \geq t_{\rm plat}$ (this effect is again seen in the numerical simulations\cite{dessart2011}). 

From the discussion above, and as we show below, we can use the observed light curves to set a robust upper limit on the value of $L_{\rm plat}t_{\rm plat}^2/2$ in the SNe in our sample (under the hypothetical assumption that they would not have contained any $\Ni$) and thus on $ET$ in that case. However, since $ET$ is independent of the amount of $\Ni$ in the ejecta, this upper limit on $ET$ is also applicable to the contribution of cooling envelope emission to the actual light curves in our sample, which are dominated by $\Ni$ power. To set this upper limit, we use the fact that, at any time, the light curve of a SN with $\Ni$ is comparable to, or brighter than, the light curve of the same explosion without $\Ni$, i.e. at all times the observed $L(t) \geq L_\text{plat}$. We therefore can treat the luminosity of the faintest data point observed before the peak in the light curve of each SNe, denoted $L_{\min}$, as an upper limit on $L_\text{plat}$. We note that the true value if $L_\text{plat}$ in most cases is likely much lower than $L_{\min}$. We know this because, in most SNe in our sample,  $L_{\min}$ is set by the first detection and, at that time, the light curve is already rising sharply, implying that $L_{\min}$ is most likely already dominated by $\Ni$ power at that stage. Taking this upper limit together with $t_p \geq t_{\rm plat}$, we obtain $L(t) t_{\rm peak}^2/2 \geq  L_{\rm plat}t_{\rm plat}^2/2 \approx ET$. Thus, if $\LTm$ in our sample is dominated by cooling emission, it must satisfy  $\LTm \lesssim L_{\rm min} t_{\rm peak}^2/2$.

Applying this test to our sample, we find that 90\% of the SNe have $\LTm$ values greater than $L_{\rm min} t_{\rm peak}^2/2$. For 19\% of the sample we even obtain $\LTm$ values 5 times greater than $L_{\rm min} t_{\rm peak}^2/2$.
These results imply that cooling emission from the explosion of a large-radius progenitor is an unlikely explanation for our excess $\LTm$ result.

\subsection*{Constraints on the central engine properties}
Here we consider the constraint on the properties of a central engine that contributes $\QTe \approx 10^{54}{-}10^{55}$\,erg\,s to the light curve of a SE~SN. We consider an engine with energy injection rise time that is shorter than its operation time and that, after its power injection peaks, its luminosity as well as $d\log(\Qe)/d\log(t)$  decline roughly monotonically. The engine can show rapid variability on timescales that are much shorter than its operation time, in which case $\Qe$ can be considered as the average energy injection over the variability time. The two main engine candidates, an accreting central compact object and a rapidly rotating magnetar, both satisfy these assumptions. Under these assumptions, $\QTe(t) \sim \Qe(t) t^2$ as long as $d\log(\Qe)/d\log(t)>-2$. Thus, defining $\te$ such that $\left. d\log(\Qe)/d\log(t)\right\rvert_{t=t_{\rm eng}} =-2$, the value of $\QTe$ at late times satisfies $\QTe(t>\te) \approx \Qe(\te) \te^2$. We define $E_{\rm eng} \approx \Qe \te$ as the energy injected over time scale $\te$. In the general case, $E_{\rm eng}$ is the total energy deposited by the engine and $\te$ is the duration of this energy injection. However, this is not the case for all types of engine temporal evolution. If there is an extended phase during which $-2<d\log(\Qe)/d\log(t)<-1$, then the total energy injected by the engine (over a time scale much shorter than $\te$) can be significantly larger than $\Ee$.  

The engine energy cannot be larger than the total explosion energy, therefore $E_{\rm eng} < E_{\rm exp}$, where $E_{\rm exp}\sim 10^{51}{-}10^{52}$\,erg. In addition, since the value of $\LTm$ is dominated by the emission during the peak, the engine energy injection must take place before or during the peak of the emission, $t_{\rm eng} \lesssim t_{\rm peak}$, where in our sample $t_{\rm peak} \sim 10^6$\,s. These upper limits together with equation~(\ref{eq:Eeng_teng}) and our finding that $\LTm \approx 10^{54}{-}10^{55}$\,erg\,s in our sample, lead to the constraints given in equation~(\ref{eq:eng_costrains}).

To conclude, the engine should inject its energy over a duration that is longer than about an hour and shorter than about a week, where the shorter time scale corresponds to an injected energy of $\sim 10^{51}{-}10^{52}$\,erg and the longer one corresponds to $\sim 10^{48}{-}10^{49}$\,erg. Next we discuss the two main candidates for central engines, accretion and a fast spinning magnetar.

\subsubsection*{Accretion onto a compact remnant}
In principle, the energy released during accretion of in-falling material onto a newly formed neutron star or black hole can provide more than enough energy to power the non-radioactive emission. Equation~(\ref{eq:Eeng_teng}) implies an accreted mass of 
\begin{equation}\label{eq:Macc}
    M_{\rm acc} \approx 10^{-4}\,{\rm M_\odot}\,\left(\frac{\epsilon_{\rm acc}}{0.1}\right)^{-1}\left(\frac{t_{\rm acc}}{10^5\,{\rm s}}\right)^{-1}\,\frac{\LTm}{10^{54}\,{\rm erg\,s}},
\end{equation}
where $t_{\rm acc}$ is the accretion time and $\epsilon_{\rm acc}$ is the efficiency at which energy is extracted from the accretion.
This is a reasonable amount of mass in the aftermath of a core collapse. The corresponding accretion rate is
\begin{equation}\label{eq:Macc}
    \dot{M}_{\rm acc} \approx 10^{-9}\,{\rm M_\odot~s^{-1}}\,\left(\frac{\epsilon_{\rm acc}}{0.1}\right)^{-1}\left(\frac{t_{\rm acc}}{10^5\,{\rm s}}\right)^{-2}\,\frac{\LTm}{10^{54}\,{\rm erg\,s}},
\end{equation}
a hyper-Eddington rate. It is much too low for efficient cooling by neutrinos, yet much higher than the $\sim 10^{-16}\,{\rm M_\odot\,s^{-1}}$ Eddington accretion rate for a disk that cools via photons.

The required accretion timescale is hard to explain naturally, where the main difficulty is how to sustain accretion for longer than $10^{3}$\,s after the ejection of the envelope. During the SN explosion $10^{51}{-}10^{52}$\,erg are deposited in the envelope. This energy unbinds the entire envelope within at most a few hundred seconds. Note that this is true even if the energy deposition is highly aspherical\cite{Eisenberg2022}. Thus, the naive expectation (see alternative option below) is that the accretion disk should form within this time, and therefore the duration of accretion that lasts for more than $10^3$\,s is most likely dominated by the accretion time of the disk, $t_\text{acc}$. This time can be estimated using an effective dimensionless viscosity parameter $\alpha$, where $t_\text{acc} \sim \alpha^{-1} T_\text{orb}/2\pi \sim 300\,\text{s}~(\alpha/0.01)^{-1} (M_{\rm rem}/M_\odot)^{1/2}(r_{\rm disk}/10^9~{\rm cm})^{3/2}$, $M_{\rm rem}$ is the compact remnant mass, $T_\text{orb}$ is the orbital time of the disk, and $r_{\rm disk}$ is its radius. Note that we assumed here a thick disk, as expected for the highly hyper-Eddington accretion rates needed to power such an engine.  Yet another constraint is on the free-fall time of the material that forms the disk, which should be shorter than the time at which the deposited energy unbinds the envelope, namely a few hundred seconds. This maximal free-fall time implies that the material that forms the disk must originate from  an initial radius of no more than about $10^{10}$\,cm in the progenitor. 

We conclude that, for a typical accretion scenario, the simple limits that we derive here allow for an accretion-powered engine, although not in a very natural way. The disk in this scenario should form at a radius of $\sim 10^{9}{-}10^{10}$\,cm, outside the radius of the pre-collapsed core. The mass from the disk should free-fall from an initial radius $\lesssim 10^{10}$\,cm, implying a fast-rotating progenitor (so that the specific angular momentum of the pre-explosion envelope fits a Keplerian orbit of the formed disk). Finally, such accretion can power only an engine with a duration that is consistent with the lower-limit of equation~(\ref{eq:eng_costrains}) ($10^3{-}10^4$\,s). 

An alternative accretion model is one where the SN shock unbinds almost the entire envelope, and sends a minute fraction  of the stellar mass (of order of $M_{\rm acc}$) to marginally bound trajectories. In such a case, the fallback of $M_{\rm acc}$ onto the compact remnant can last as long as needed to power the engine. Testing this model is extremely difficult, as it is challenging to follow numerically the hydrodynamics of such a small fraction of the ejecta mass.

\subsubsection*{Magnetar properties}
If the engine is a magnetar, then we can further constrain its magnetic field and initial rotation. The energy that can be contributed by a magnetar is its rotational energy, $E_\text{mag}\approx 2 \times 10^{52}\,{\rm erg}\,P_{i,\text{ms}}^{-2}$
where $P_i=P_{i,\text{ms}}\cdot 1$\,ms is its initial rotation period. Dividing this energy by the spin-down luminosity\citep{spitkovsky2006}, one obtains the time over which this energy is injected into the ejecta, $t_\text{mag} \approx 10^3\,{\rm s}\,P_{i,\text{ms}}^{2}\,B_{15}^{-2}$
where $B=B_{15}\times 10^{15}$\,G is the surface dipolar magnetic field. The luminosity of the magnetar wind is roughly constant over $t_\text{mag}$, so its value is about $E_\text{mag}/t_\text{mag}$ during this time. At $t>t_\text{mag}$ the luminosity drops as $t^{-2}$, so under the assumption that the entire wind energy thermalizes, $\QTe(t)$ increases logarithmically at $t>t_\text{mag}$, and therefore $\QTe(t>t_\text{mag})  \approx E_\text{mag} t_\text{mag} \ln(t/t_\text{mag})$. Thus, requiring $\QTe(t>t_{\rm peak}) \approx \LTm$, we obtain the constraints on the magnetar dipolar magnetic field (equation~\ref{eq:Bmag}) and on the initial spin period.

\bmhead{Acknowledgements}
We thank Amir Sharon,  Doron Kushnir and Avishai Gilkis for comments and discussions. 
This work was supported by grants No. 818899 (E.N.) and 833031 (D.M.) from the European Research Council (ERC).

\bmhead{Author Contributions}
O.R and E.N. conceived the idea to measure the Katz integral for these data. O.R. performed the data analysis, revealing the excess power source. E.N. led the theoretical analysis of the nature of the 
excess. 
All three authors took part in the discussions, analysis, and writing of the manuscript. Correspondence and requests for materials should be addressed to O.R and/or E.N.
\bmhead{Competing interests} The authors declare no competing interests.

\bmhead{Data availability}
The data used to create the figures in this study, as well as the
scripts used to generate the figures, are available online in Zenodo with the identifier

\bmhead{Code availability}
We used Python 3.7.6 and the public Python packages NumPy 1.19.4 and Matplotlib 3.4.1. The stellar-evolution code MESA is freely available and documented at \href{https://docs.mesastar.org/en/release-r23.05.1/}{https://docs.mesastar.org/en/release-r23.05.1/}.

\makeatletter
\apptocmd{\thebibliography}{\global\c@NAT@ctr 27\relax}{}{}
\makeatother

\newpage

\begin{table}
\caption{SN sample.}
\label{table:SN_sample}
\includegraphics[width=1.0\columnwidth]{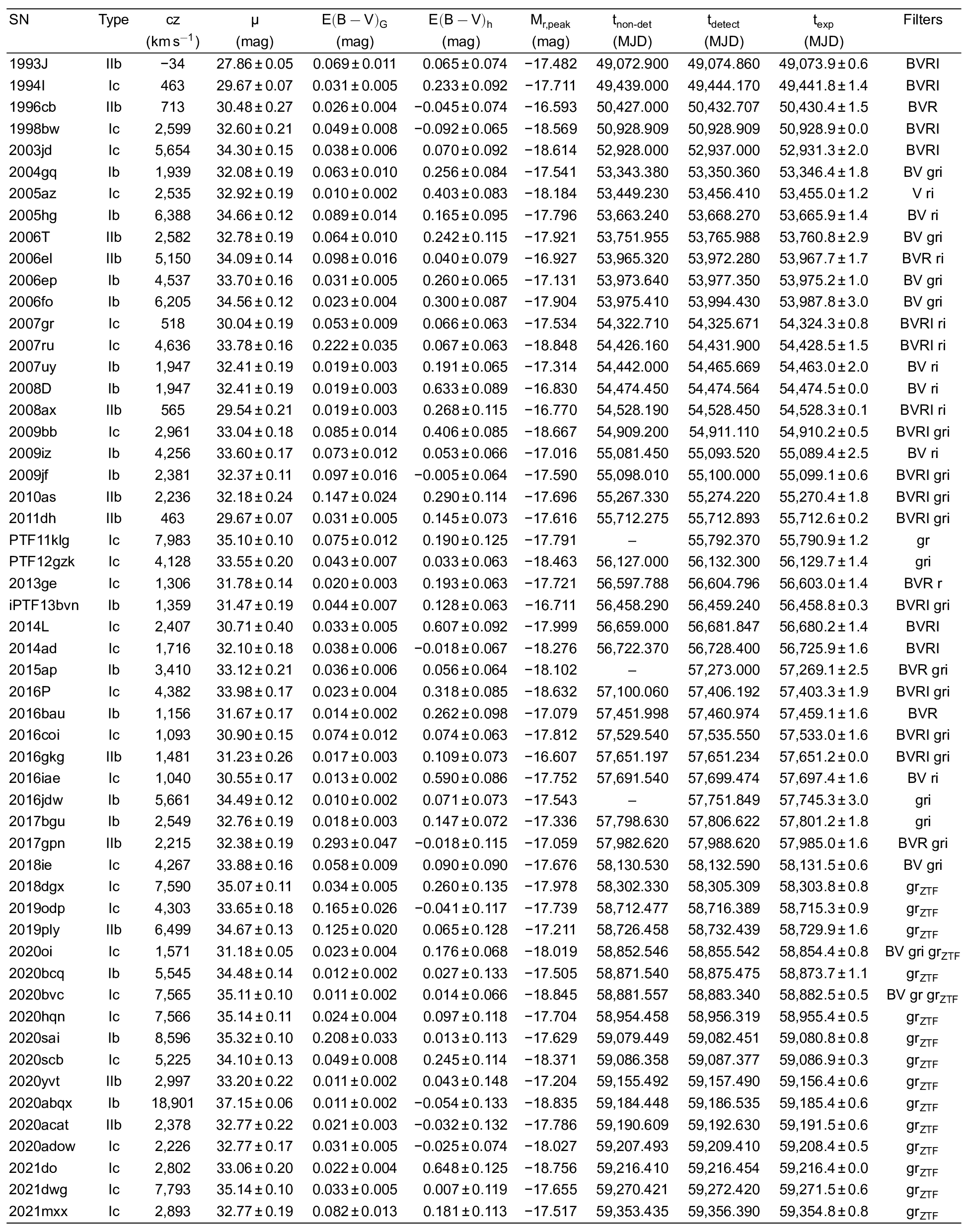} 
Uncertainties are $1\,\sigma$ errors.
\end{table}

\begin{table}
\caption{SN parameters.}
\label{table:SN_parameters}
\includegraphics[width=1.0\columnwidth]{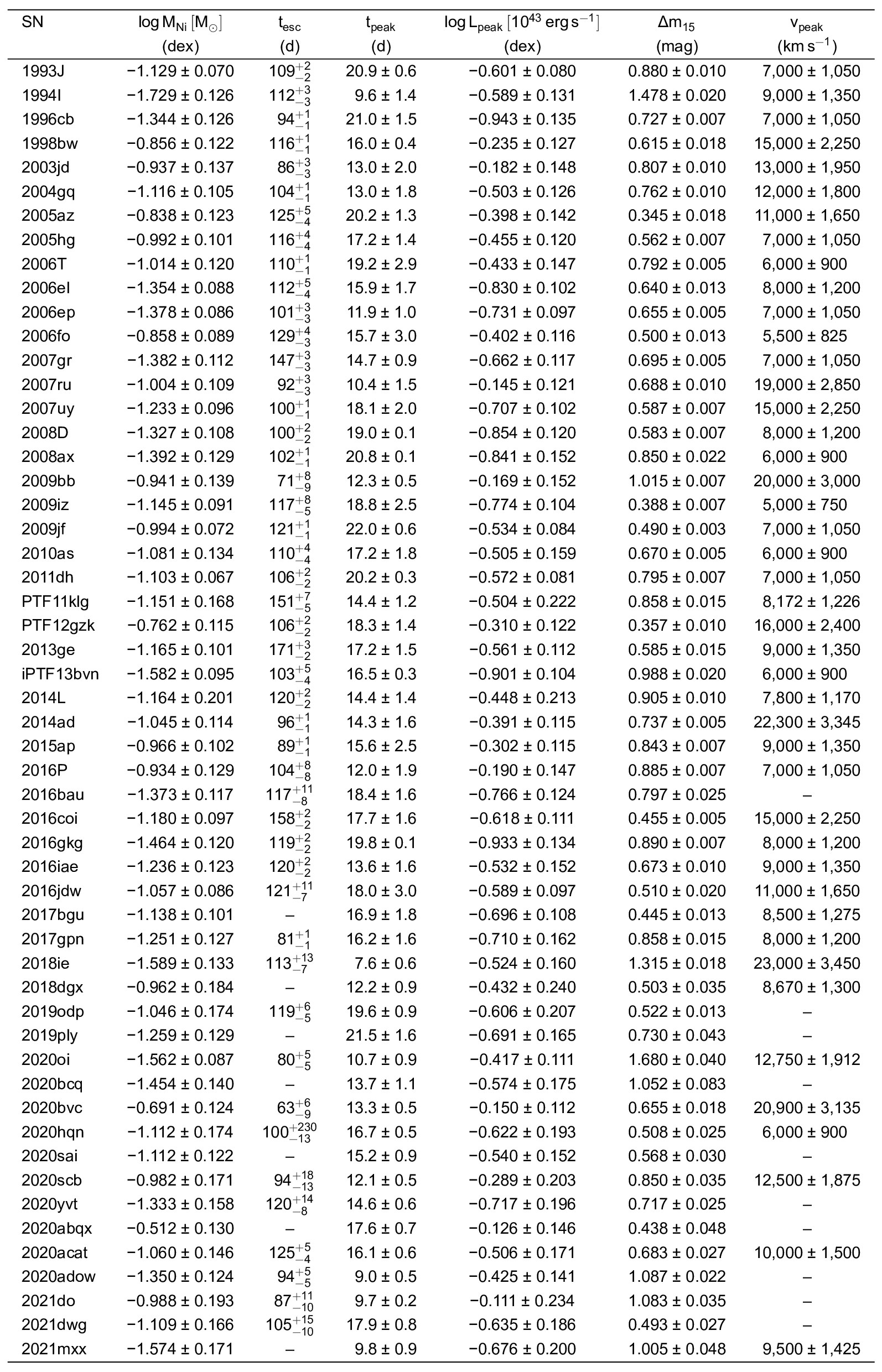}
Uncertainties are $1\,\sigma$ errors, while lower and upper $t_\text{esc}$ errors are 16th and 84th percentiles.
\end{table}

\begin{table}
\caption{$\LTm$ and $\LTm/LT_{100}$ values.}
\label{table:LT-nuc}
\includegraphics[width=1.0\columnwidth]{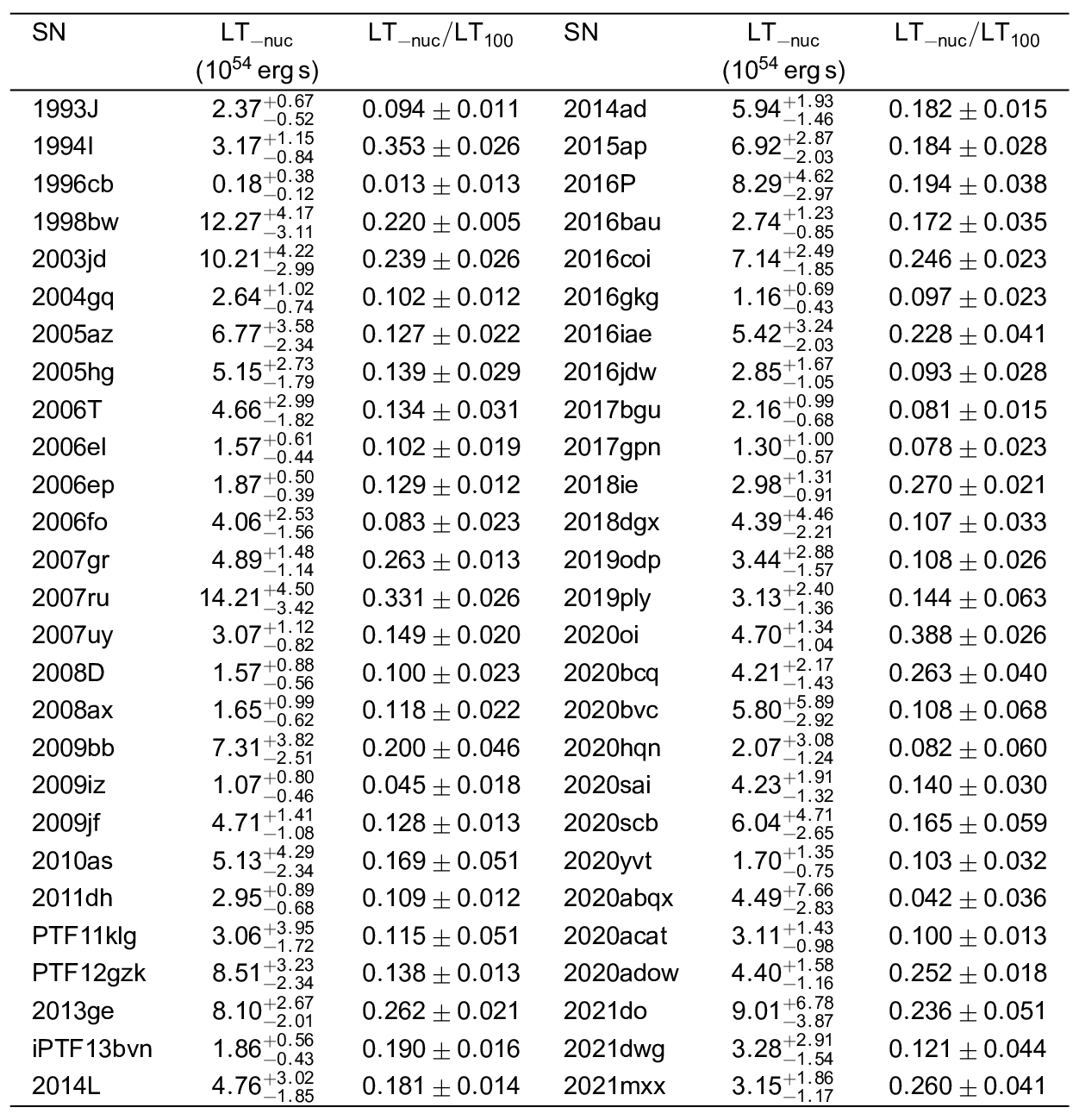}
Lower and upper $\LTm$ errors are 16th and 84th percentiles, while $\LTm/LT_{100}$ uncertainties are $1\,\sigma$ errors.
\end{table}

\end{document}